%% file: Arxiv_Transformer_aug_2024_one_column.tex
\documentclass[lettersize,journal,onecolumn,draftclsnofoot]{IEEEtran}
\usepackage{amsmath,amsfonts}
\usepackage{array}
\usepackage[font=footnotesize]{caption}
\usepackage{subcaption}

\usepackage{epsfig}
\usepackage{algorithm}
\usepackage{algpseudocode}
\usepackage{placeins}

\usepackage{textcomp}
\usepackage{stfloats}
\usepackage{url}
\usepackage{verbatim}
\usepackage{cite}
\usepackage{todonotes}
\usepackage{float}
\usepackage{varwidth}
\usepackage{makecell} 
\usepackage{multirow}
\usepackage{bm}

\algrenewcommand\algorithmicrequire{\textbf{Input:}}
\algrenewcommand\algorithmicensure{\textbf{Output:}}

\begin{document}

\title{Application of Transformers for Nonlinear Channel Compensation in Optical Systems}

\author{Behnam Behinaein Hamgini, Hossein Najafi, Ali Bakhshali, and Zhuhong Zhang
\thanks{Authors are with Huawei Technologies Canada, Ottawa R\&D Centre, 303 Terry Fox Drive,Kanata, 
Ontario K2K 3J1.\\
Emails: behnam.behinaein.hamgini, hossein.najafi,  ali.bakhshali, and zhuhong.zhang@huawei.com.
}
}



\maketitle

\begin{abstract}
In this paper, we introduce a new nonlinear optical channel equalizer based on Transformers.
By leveraging parallel computation and attending directly to the memory across a sequence of symbols, we show that Transformers can be used effectively for nonlinear compensation (NLC) in coherent long-haul transmission systems.
For this application, we present an implementation of the encoder part of the Transformer and analyze its performance over a wide range of different hyper-parameters. 
It is shown that by proper embeddings and processing blocks of symbols at each iteration and also carefully selecting subsets of the encoder's output to be processed together, an efficient nonlinear equalization can be achieved for different complexity constraints. 
To reduce the computational complexity of the attention mechanism, we further propose the use of a physic-informed mask inspired by nonlinear perturbation theory. 
We also compare the Transformer-NLC with digital back-propagation (DBP) under different transmission scenarios in order to demonstrate the flexibility and generalizability of the proposed data-driven solution.

\end{abstract}


\section{Introduction}\label{intro}
\IEEEPARstart{T}{he} impact of nonlinear interference from the fiber Kerr effect and optical/electrical components is a main challenge for high-speed optical systems that limits the achievable transmission rates. 
In general, fiber nonlinearity interference can be equalized in sample or symbol domain by approximating and inversing the nonlinear Schrodinger equation through deterministic models such as DBP \cite{DBP1, DBP2, DBP3} or perturbation-based nonlinear compensation (PNLC) methods \cite{pert1, pnlc3}. 
Alternatively, various solutions have been proposed in order to exploit artificial neural networks (ANN) for nonlinear compensation in optical communications \cite{NLC-MLP1, hager2018nonlinear, CNN1}.

The pattern- and medium-dependent characteristics of nonlinear propagation make it a suitable case for artificial neural network domain. 
Specifically, recurrent neural networks (RNN) such as long short-term memory (LSTM) 
operating on received symbol sequences have shown to be effective in fiber nonlinearity compensation for different transmission scenarios.
\cite{deligiannidis2020compensation,perf-comp1, Bakhshali2023NeuralDMB}.
Attempts are also made to design appropriate ANNs for hardware implementation 
with real-world applications in mind \cite{co-lstm, freire2023reducing_complex_lstm}.
However, the sequential nature of RNNs at the core of better performing architectures, is a bottleneck for parallel implantation that is vital for hardware developments of ultra-high-speed optical transmission systems.
On the other hand, Transformers as powerful ANN structures based on self-attention 
mechanism have been introduced in \cite{vaswani2017attention}.
Since their birth, these structures are widely employed in various machine learning applications with impressive performance. 
In fact, Transformers are designed to learn long memories while overcoming the limits of sequential nature of RNNs. In particular, unlike RNNs, Transformers are highly parallelizable in both training and inference stages which makes them greatly suitable for hardware developments of ultra-high-speed optical transmissions.

In this work, we employ Transformers for applications in coherent optical communications.
To the best of our knowledge for the first time, we study the application of Transformer as the core of fiber nonlinear equalization and explore various opportunities and challenges associated to this architecture.
We show that in order to leverage parallel computation and attending directly to the memory across a sequence of received symbols, Transformers with efficient embeddings and implementation can be used for nonlinear compensation.
One major advantage of Transformers over LSTMs for nonlinear compensation application is that both the training and inference are performed over a block of received symbols. RNNs  generally suffer from the vanishing gradient during the training stage as well as increased latency in the inference stage over long blocks of symbols \cite{co-lstm}. In contrast, Transformer-NLC can be directly optimized over the target block length and is inherently more suitable for applications in high-speed coherent optical transceivers that require a great deal of parallelization to satisfy their high-throughput and low-latency requirements.

We also present the use of a physic-informed mask according to the nonlinear perturbation theory in order to make the attention matrix sparse. 
A two-dimensional relation, modeled as a mask matrix, highlights the importance of combinations of symbols for the nonlinear estimation and is shown to generally provide savings in the computational complexity while particularly demonstrates performance advantage at regions with limited resources.
Finally, we present a comparison of performance and complexity of Transformers with DBP for nonlinear channel compensation under different transmission scenarios and demonstrate the capabilities of Transformers compared to DBP as a flexible universal solution for optical networks. 
\IEEEpubidadjcol

The remainder of this paper is organized as follows: In Section \ref{TF_Pre}, the structure of fiber nonlinear compensation is briefly discussed and the Transformer's basics are reviewed. 
In Section \ref{TF_main}, the Transformer-NLC architecture is introduced and then the impact of perturbation mask for reducing the computational complexity of the attention mechanism is explained. 
Next, numerical results are presented in Section \ref{TF_sim} and finally, we summarize the paper in Section \ref{TF_conc}.

\section{Preliminaries}\label{TF_Pre}
\subsection{Nonlinear Compensation in Coherent Optical Systems}

Considering the dual-polarization evolution of optical field over a fiber link according to the Manakov equation \cite{Manakov}, the nonlinear propagation impact can be shown as the right-hand side of the following equation:
\begin{align}
		\frac{\partial{u_{x/y}}}{\partial{z}} + \frac{\alpha}{2}u_{x/y} + j\frac{\beta}{2}\frac{\partial^2u_{x/y}}{\partial{t^2}} =
		 j\frac{8}{9}\gamma\Bigl[|u_x|^2+|u_y|^2\Bigr]u_{x/y},
		\label{equ:manakov}
\end{align}
where $u_{x/y} = u_{x/y}(t,z)$ represents the optical field of polarization $x$ and $y$, respectively, $\alpha$ is the attenuation coefficient, $\beta$ is the group velocity dispersion (GVD), and $\gamma$ is the nonlinear coefficient.
Kerr nonlinear impact can traditionally be equalized through deterministic nonlinear models by approximating and inversing the Eq.~\eqref{equ:manakov} through DBP \cite{DBP1, DBP2, DBP3} where the fiber is modeled as a series of linear and nonlinear sections through first-order approximation.
Also, by employing the perturbation analysis \cite{pert1}, the optical field can be represented as the solution of a linear term plus a nonlinear perturbation term in symbol domain as a lumped stage. In fact, the first-order perturbation term can be modeled as the weighted sum of triplets of symbols in addition to a constant phase rotation\cite{pnlc3, pnlc4}.

\begin{figure}[b]
	\centering	
	\includegraphics[width=0.65\linewidth]{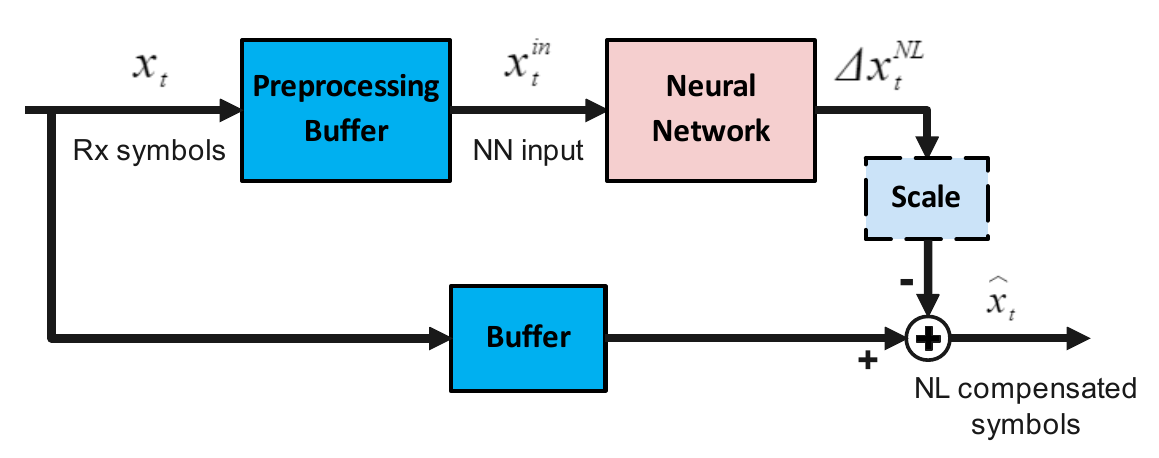}
	\caption{An example of receiver symbol-domain nonlinear compensation with neural network.}
	\label{diag:ANN}
\end{figure}

Applying data-driven models, various solutions are introduced to employ machine learning for the fiber nonlinearity compensation(e.g. \cite{NLC-MLP1, hager2018nonlinear, CNN1, deligiannidis2020compensation, perf-comp1, Bakhshali2023NeuralDMB}) in which one can learn the equalization process through data. 
These artificial neural network nonlinear compensation (ANN-NLC) solutions include learned-DBP, using triplets of symbols for PNLC with basic neural networks, and using soft-symbols in deep neural networks.
Basically, a data-driven ANN solution provides a more general and flexible approach while providing the potentials for large simplifications in computational complexity with efficient pruning and quantization of the network \cite{QNN1, gholami2022survey, qin2020binary, hamgini2024pruning}.

Considering one-stage nonlinear compensation, a basic block diagram for the receiver side equalization process is presented in Fig.~\ref{diag:ANN} where after the conventional linear processing for the coherent optical receiver, the preprocessing buffer generates appropriate inputs for an ANN method. 
This can be seen in form of calculated PNLC triplets for a perturbation-based method, or a vector of soft symbols at the end of Rx-DSP chain for a deep artificial neural network nonlinear compensation.
Assuming the first-order perturbation is the dominant term, a scaler can be used to adjust the model for different optical launch powers:
\begin{equation} 
	\alpha = 10^{\bigl(P_\text{inference}(\text{dB}) - 
		P_\text{train}(\text{dB})\bigr)/10},
	\label{ANN_scale}
\end{equation}
where $P_\text{train}$ is the optical launch power of the data used in the training while $P_\text{inference}$ is the respective optical launch power of data in the inference (equalization) stage.
This in fact enables a universal model that can be trained and deployed at various launch powers. Moreover, to handle other changes in a transmission scenario, transfer learning can be employed which updates only a part of the ANN \cite{transfer_learn}.

\subsection{Transformers}\label{TF_basics}
Transformers were introduced in \cite{vaswani2017attention} based on the self-attention mechanism. They have been employed in different fields of machine learning including language modeling, vision, audio, and time series processing with impressive performance while being appropriate for parallel implementation. 
Specifically, a Transformer processes all symbols in parallel and provides direct interactions among symbols in the input sequence which makes it capable of capturing memories more efficiently compared to an LSTM where the memory is handled sequentially. 
This makes Transformers highly matched for real-time applications such as hardware developments of ultra-high-speed optical transmissions.

\begin{figure}[tb] 
	\centering
	\includegraphics[clip,width=0.8\linewidth]{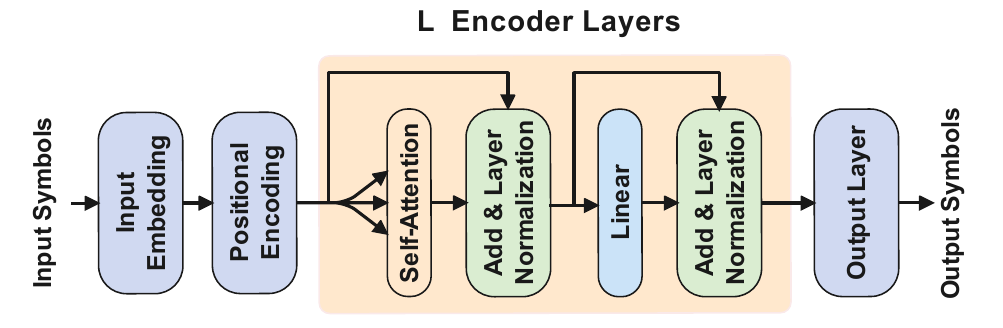}
	\caption{ Transformer encoder architecture.}
	\label{fig:transformerencoder}
\end{figure}

A vanilla Transformer, originally developed for natural language processing, consists of an encoder-decoder architecture.
However, for our application of nonlinear equalization, we just employ the encoder part of a Transformer which is depicted in Fig.~\ref{fig:transformerencoder}.
This architecture consists of a module that generates the embeddings, a positional encoder, $L$ stacked layers of multi-head attention (MHA), position-wise feed-forward network (FFN), and an output module at end that can generate the estimated nonlinear interference associated to real and imaginary parts of the equalized output symbol sequence.
Furthermore, there are residual connections and a layer normalization at each Transformer layer. 

The attention function is at the core of a Transformer where a query-key-value vector model is employed. 
Attention can be computed using various methods including, additive attention \cite{bahdanau2014neural}, dot-product attention \cite{vaswani2017attention}, kernel-based attention \cite{katharopoulos2020Trafsformers_R_RRN}, or attention through a set of 
learned position biases \cite{zhai2021attention-free}.
Specifically for the scaled dot-product self-attention, given queries $ Q \in R^{N \times d_K} $, keys $ K \in  R^{M \times d_K}$, and values $ V \in  R^{M \times d_V} $, one can write: 
\begin{align} \label{equ:self_attn}
	&A = \text{softmax}(\frac{QK^T}{\sqrt{d_K}}) \\ \nonumber
	&\text{Attention}(Q, K, V) =  AV
\end{align}
\noindent where $ N $ and $ M $ are the length of queries and keys (values), 
respectively, $ d_K $ is the dimension of keys and queries, $ d_V $ is the 
dimension of values, and $ A $ is the \emph{attention matrix}. 
It should be noted that the softmax is preformed row-wise in Eq.~\eqref{equ:self_attn}. 
The term $ \sqrt{d_k} $ is also added to mitigate the gradient vanishing problem of the softmax function.
In the encoder architecture, queries, keys, and values are obtained from the same input sequence.
Moreover, a vector of learned embeddings can be used instead of original input sequence to the encoder since the embedding generating module generally produces better input features. 

Transformer architecture utilizes a multi-head attention mechanism to map the original queries, keys, and values ($ d_{model} $) into lower-dimension spaces, $ d_K, d_V $ using linear layers where $ d_K = d_V = d_{model} / h $, and $ h $ is the number of heads. After applying the self-attention showed in Eq. \eqref{equ:self_attn}, the output of multi-head attentions are concatenated, passed through a linear layer, and fed into the position-wise FFN module. 
The FFN is a fully connected feed-forward module located at every layer of the encoder where it is applied to each position, similarly. 
This module consists of two linear layers and a ReLU activation as
\begin{equation}
	\text{FNN}(x) = \text{ReLU}(xW_1+b_1)W_2+b_2
\end{equation}
where $ W_1 \in R^ {d_{model}\times d_{ff}} $, $ W_2 \in R^ {d_{ff}\times d_{model}} $, $ b_1 \in R^ {d_{ff}} $, and $ b_2 \in R^ {d_{model}} $.

Transformers also use residual connections and layer normalization modules after each pair of attention and FNN modules as described by:
\begin{eqnarray}
	Y = \text{LayerNorm}(\text{MHA}(X) + X) \\ \nonumber
	Z = \text{LayerNorm}(\text{FFN}(Y) + Y),
\end{eqnarray}
where the layer normalization is performed after the residual addition. Note that changing the layer normalization position in the network leads to different Transformer structures. This may impact the training behavior and learning rate optimization\cite{xiong2020Pre-LN-TF}. 
However, in this work we use the original post layer normalization structure which needs a warm-up stage for the learning rate.
Furthermore, we use a positional encoder as introduced in \cite{vaswani2017attention}.

\section{Transformers for Nonlinear Compensation}\label{TF_main}

\subsection{Model Architecture}\label{TF_overview}
\begin{figure}[t] 
	\centering
	\includegraphics[clip, width=0.8\linewidth]{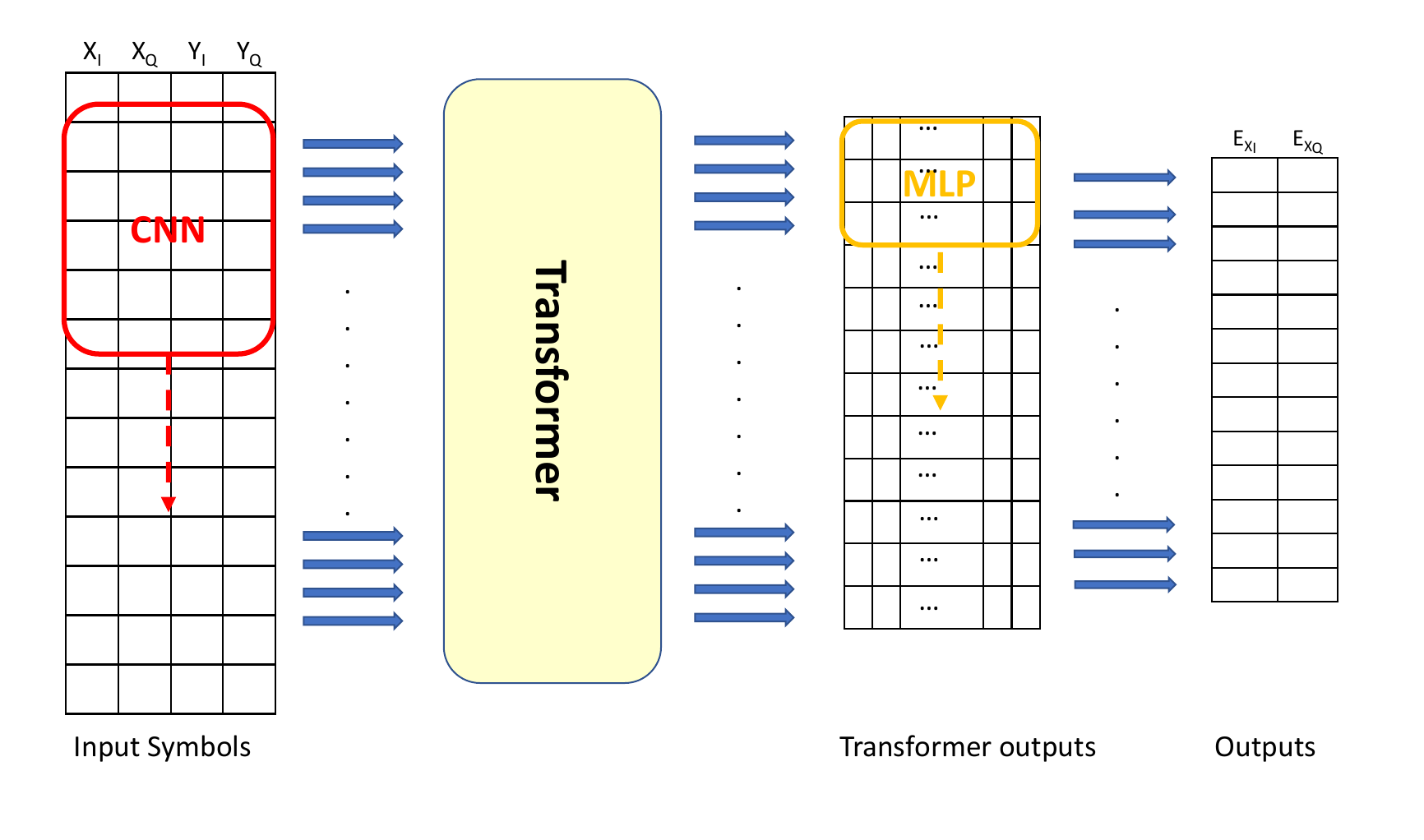}
	\caption{The overall architecture for the nonlinear channel equalization. From left to right: the CNN generates input embeddings which are processed by the Transformer to generate output representations. These outputs are then fed into an MLP to generate the estimated nonlinear distortions $E_{XI}$ and $E_{XQ}$.}
	\label{fig:overall_archit}
\end{figure}

In this work, we use the above Transformer encoder architecture to estimate the nonlinear distortions in the received symbols. 
The overall design for the nonlinear channel equalization is depicted in Fig.~\ref{fig:overall_archit}. 
%
Due to the channel memory and pattern-dependency over the transmitted sequence, it is required to process several symbols before and after each target symbol. 
The number of these extra symbols from each side of the target symbol is denoted by tap-size value $t$. 
%
%
The model uses received symbols from four components of the optical domain at the end of the DSP chain, $X_I, X_Q, Y_I, Y_Q$. 
At first, an embedding generator module maps this sequence of received symbols into $ R^{N \times d_{model}} $, where $N$ is the embedding sequence length.
%
%
%
Next, a Transformer is used as the core for nonlinear channel equalization. 
The dashed block in Fig.~\ref{fig:transformerencoder} corresponds to the Transformer block that operates on the generated embeddings.
The last module processes the Transformer's output to generate the equalizer's nonlinear distortion estimations for one polarization (or for both polarizations by slightly modifying this output module).

Considering the ultra high-speed optical applications, parallel and vector processing (instead of sequentially processing of a single symbol) is vital to achieve high throughput out of the digital processing stage. Therefore, it is required to employ the NLC equalizer over blocks of data in the inference stage.
%
%
Transformers in contrast to LSTMs, have the benefit of being trainable on very long sequences of symbols without the vanishing gradient issue or delay of sequential processing, which yields the advantage of accelerating the training process by a large factor. 
Here, we complete the Transformer-NLC design by employing a sequence of symbols to produce a series of associated nonlinear distortions at the output, concurrently.
%

For processing a block of symbols for equalization, the training/inference samples are created by selecting a block of $b$ symbols. 
In order to maintain the quality of equalization at the beginning and end of the block, we append extra symbols to the beginning and end of the block; effectively expanding the selected symbols with extra $t$ symbols at each side. Hence, each input sequence length is equal to $ b + 2t$. 
This sequence of symbols passes through the embedding generator module which produces $ b + \ell$ embeddings (the exact value of $\ell$ will be explained later). These embeddings are then fed into the Transformer where $N = b + \ell$ outputs are generated.

The output module (MLP in the Fig. \ref{fig:overall_archit}) generates $b$ estimated nonlinear distortions which are used for computing the loss in the training stage. This loss is generated as the mean of squared differences between $b$ estimated nonlinear distortions and $b$ target ones. Through back-propagation, the optimizer uses this loss to modify the weights in the training stage.
With this approach, training can be accelerated by a large factor since instead of training over one symbol at a time, the model is trained on $b$ symbols at each training iteration. 
In addition, in the inference (equalization) stage, the Transformer generates $b$ nonlinear distortions at once through operations that are highly parallelizable.
%
Block processing reduces the computational complexity of inference which
can be seen from the following equation where the computational complexity of the attention matrix is expressed as a function of block size $b$:
\begin{equation}\label{equ:mha_complexity_block}
	RMPS_{att-blk} = \frac{2h(b+\ell)^2d_k + 3h(b+\ell)^2}{b}.
\end{equation} 
where $h$ is the number of heads and RMPS stands for real multiplications per symbol. In analyzing the computational complexity, we only consider the multiplication operations and do not include the additions.  Hence, RMPS is used as the complexity metric in this work. The first term in the above equation is the complexity of matrix multiplications 
and the second term is the complexity of division by  $ \sqrt{d_k} $ and  softmax computation. In a vanilla Transformer, the size of $ d_K $ and $ d_V $ are chosen as $ d_{model} / h $. 
In our studies, we also explore the configurations where  $ d_K $ and $ d_V $ are equal to $ d / h $ where  $ d$ may not be equal to $d_{model}$. 
This case is particularly of interest since for $ d < d_{model} $, we can reduce the computational complexity. Note that, ReLU and leaky ReLU's complexities are excluded in our analysis, since their hardware implementations are fairly simple and requires very little resources. 
Note that the reduction factor of $b$ also exists in the other layers including the FFNs and output layer.
Assuming $ \ell $ and $h$  as constants, for large $b$, the overall complexity in Eq.~\eqref{equ:mha_complexity_block} is approximately $O(bd_{k})$ per symbol which is in fact linear in block size $ b $.

\subsection{Embedding}\label{subsec:embedding}
Our studies over the course of this research show that the Transformer which directly operates on input symbols cannot effectively learn the nonlinear interactions among the symbols. Hence, we need to create good representations, commonly known as \emph{embeddings}, of the input components, $X_I, X_Q, Y_I, Y_Q$, prior to feeding them to the Transformer.
Here, we explore two main classes of ANN models for generating the embeddings, an MLP and a CNN. 
For the MLP architecture, we examine two configurations. In the first one, an MLP with only one linear layer is used. In the second one, we employ a multi-layer MLP comprising a linear layer, a Leaky ReLU nonlinear activation with the negative slope of $0.2$, and another linear layer. 
Using these MLPs, we project the input symbols ($ R^{(2t+b) \times 4} $) into embeddings with dimension $ R^{(\ell+b) \times d_{model}} $ where $b$ is the processing block size, $ \ell = 2t $ and $t$ is the tap size as defined before. 
Note that the MLPs are applied similarly for each symbol over the input sequence.

We also explore the use of CNNs for creating embeddings since they have shown great potential for generating features. 
We use a CNN with one layer of one-dimensional convolutional layer with a kernel size of $ k $, four input channels, $ d_{model} $ output channels, and a stride of one, followed by a leaky ReLU nonlinear activation function with negative slope of $0.2$. 
The CNN generates embeddings of dimension $ R^{(\ell + b) \times d_{model}} $ where $ \ell = 2t - k + 1$.  
Note that we use $\ell$ to unify our math notation for representing computation complexities for Transformers with either MLP or CNN embeddings-generator modules. In other words, $\ell$ is equal to $2t$ for MLP generated embeddings and $2t-k+1$ for CNN ones. 

\subsection{Mask} \label{sec:mask}
The attention matrix can be \emph{masked} for variety of reasons, for example to introduce some properties into the model such as preserving the auto-regressive property. Masking can also be used to reduce the complexity of attention calculations ($A$ in Eq.~ \eqref{equ:self_attn}). 
One approach is to make the attention matrix sparse so that the zero elements are not computed or stored \cite{child2019generating, beltagy2020longformer} since computing attention is a complexity bottleneck in the vanilla Transformer for long sequences.
In this work, we propose a method to make the matrix $ A $ sparse according to a physic-informed mask from the perturbation theory. 
As briefly discussed in Section \ref{TF_Pre}, the optical field can be represented as the solution of a linear term plus a nonlinear perturbation term in symbol domain. The perturbed term can be represented as the weighted sum of symbol triplets assuming pulse spreading (memory) that is much greater than the symbol duration due to the accumulated chromatic dispersion in the fiber \cite{pnlc3}. The perturbation term for the target symbol at the middle of sequence can be represented as:
\begin{equation}\label{equ:triplets}
		\Delta u_{x/y} = \sum_{m, n} P_0^{3/2} (A_{n,x/y} A^*_{m+n,x/y} A_{m,x/y} 
		+ A_{n,y/x} A^*_{m+n,y/x} A_{m,x/y}) C_{m,n}
\end{equation}
where $P_0$ is the optical launch power, $A_{.,x/y}$ is the sequence of complex transmitted symbols, and $ C_{m, n} $ is the nonlinear perturbation coefficient.
Specifically, the perturbation coefficients $C_{m,n}$ show which combinations of symbols are more important to estimate the nonlinear interference of the target symbol.
Therefore, we can use this two-dimensional relations inside our attention matrix as a mask to reduce the computational complexity.

Considering the hyperbolic form of the perturbation coefficients, we can use a two-dimensional link-independent criteria as a mask by selecting a subset of elements in attention matrix. 
As an instance, we use the following relation for the numerical result as in \cite{NLC-MLP1}:
\begin{equation}\label{equ:mnselection}
	|n| \le \min \left( \frac{\rho \lceil \ell/2 \rceil}{|m|}, \left \lceil \frac{\ell}{2} \right \rceil \right) 
\end{equation} 
where $ \lceil \cdot \rceil $ represents the ceiling function, $ \ell $ is the maximum of $ m $ and $ n $, and $ \rho $ restrains the maximum of $ (m,n)$ pairs. 
Note that $ \rho $ and $\ell$ (which is related to the tap-size $t$) should be optimized according to the transmission scenario.

Since the Transformer also needs to determine the interaction between symbols, based on the above equation, we propose only to use the elements of attention matrix $ A_{m,n} $ that satisfy the $ m $ and $ n $'s relation in Eq. \eqref{equ:mnselection}, where $m$ and $n$ are row and column indices, respectively, and make all the other elements zeros. 
To implement this approach in the individual symbol processing, in the simulation, we add a mask matrix, $M^{indv}$ to $ QK^T/\sqrt{d_k} $. The matrix, $M^{indv}$, has the same size as $ QK^T$ and is generated such that the elements that satisfy the relationship between $m$ and $n$ in Eq. \eqref{equ:mnselection} are filled by zeros and those that do not are filled with the negative infinity. 
This guaranties that the elements in attention matrix which do not satisfy Eq. \eqref{equ:mnselection} are set to zeros since the softmax operator maps the negative infinity to zero. The elements that their indices meet Eq. \eqref{equ:mnselection} are not affected by the mask since adding a zero to an element has no effects on the softmax's output. 
In summary, with the assumption that computation of the attention matrix can be implemented such that the computations are only carried out for the non-zero values in the attention matrix (corresponding to zero values in the proposed mask), we can reduce the computational/space complexity of attention to $ O(\hat{N}d_k) $ where $ \hat{N} $ is the number of non-zero attention elements. 

\begin{figure*}[t]
	\centering
	\begin{subfigure}{0.24\linewidth}
		\centering
		\includegraphics[width=.78\linewidth]{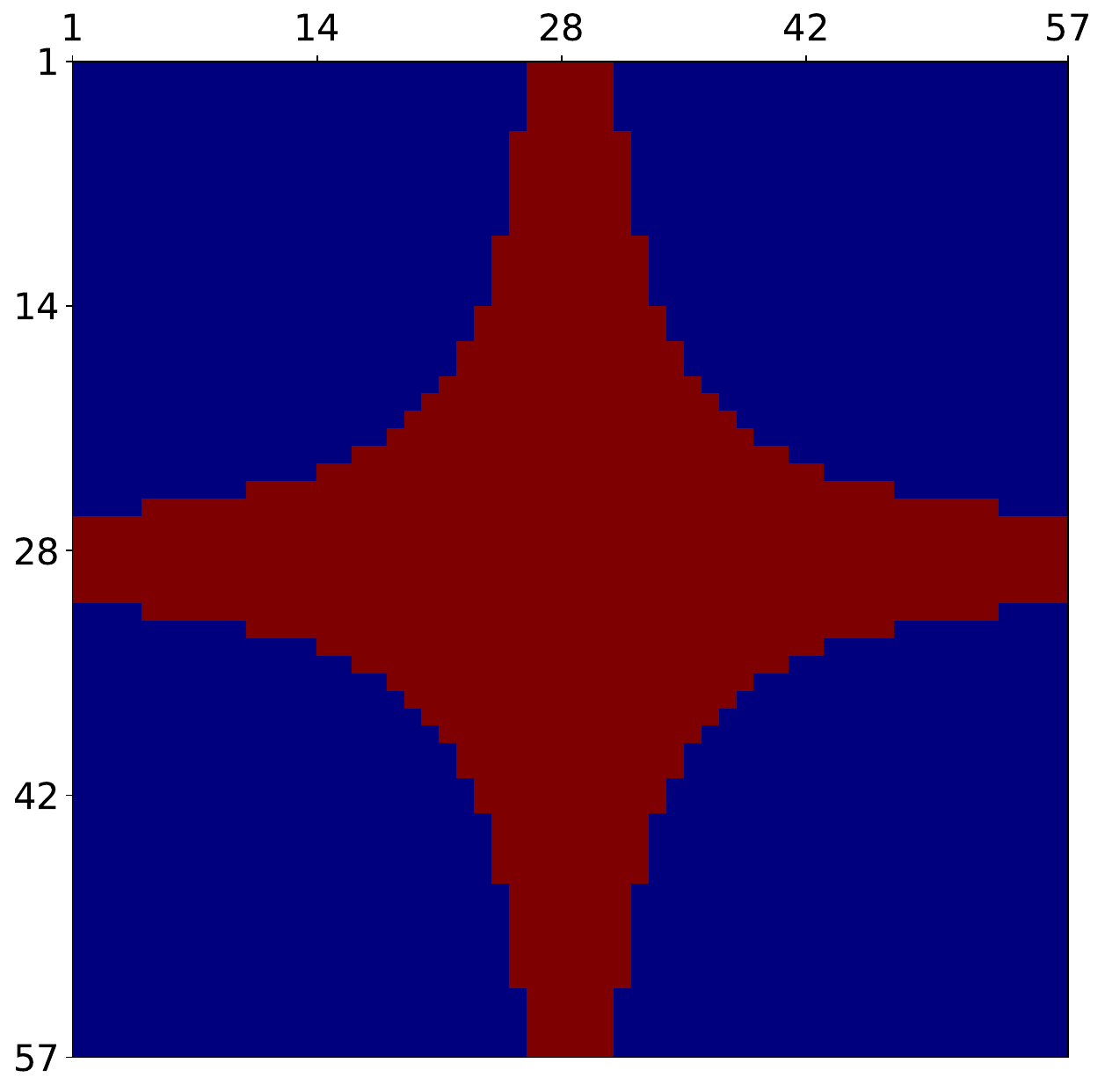}
		\caption{}
		\label{fig:attentionmaskblock1rho26t32}
	\end{subfigure}
	\begin{subfigure}{0.24\linewidth}
		\centering
		\includegraphics[width=.8\linewidth]{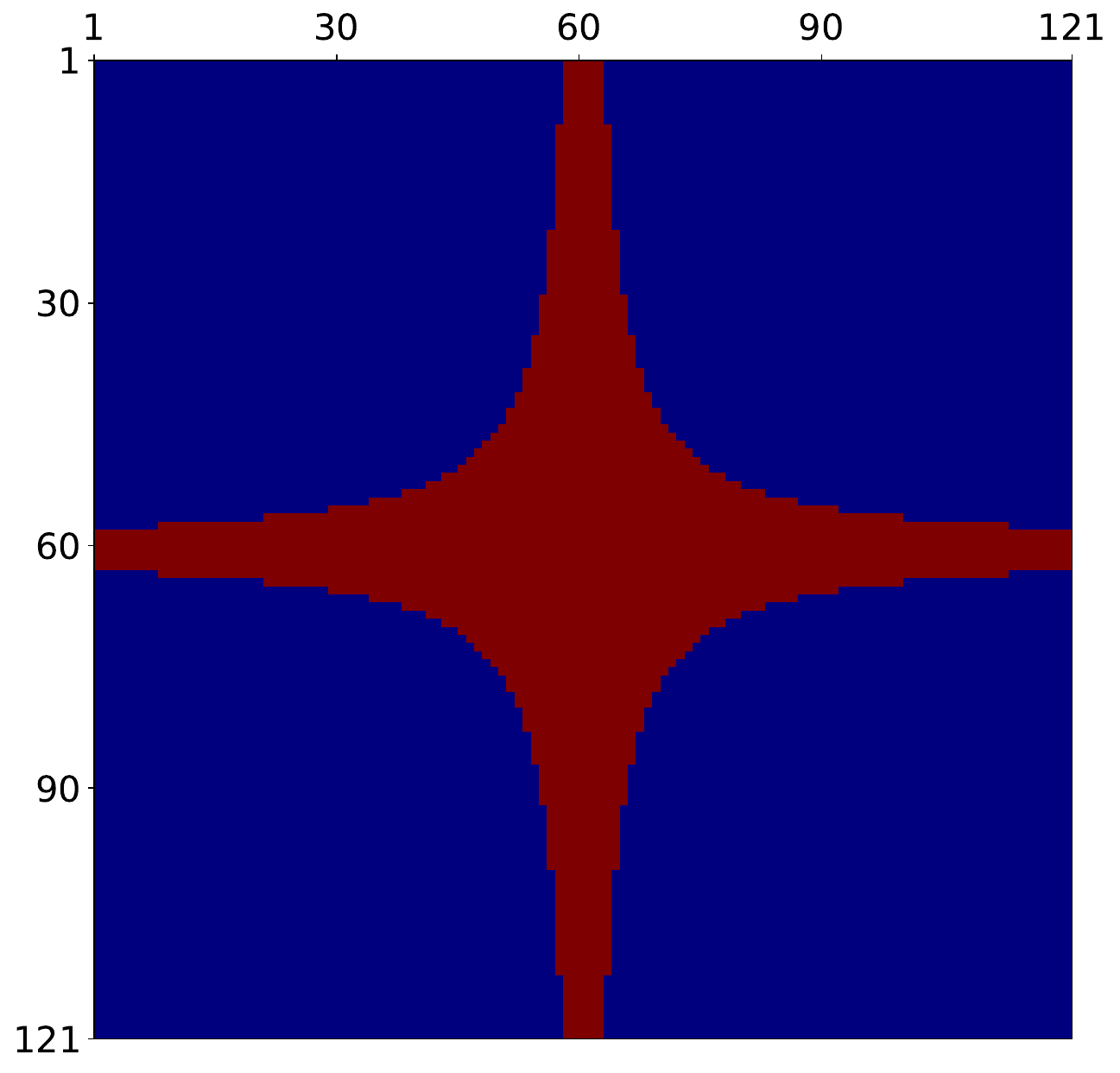}
		\caption{}
		\label{fig:attentionmaskblock1rho26t64}
	\end{subfigure}
	\begin{subfigure}{0.24\linewidth}
		\centering
		\includegraphics[width=.8\linewidth]{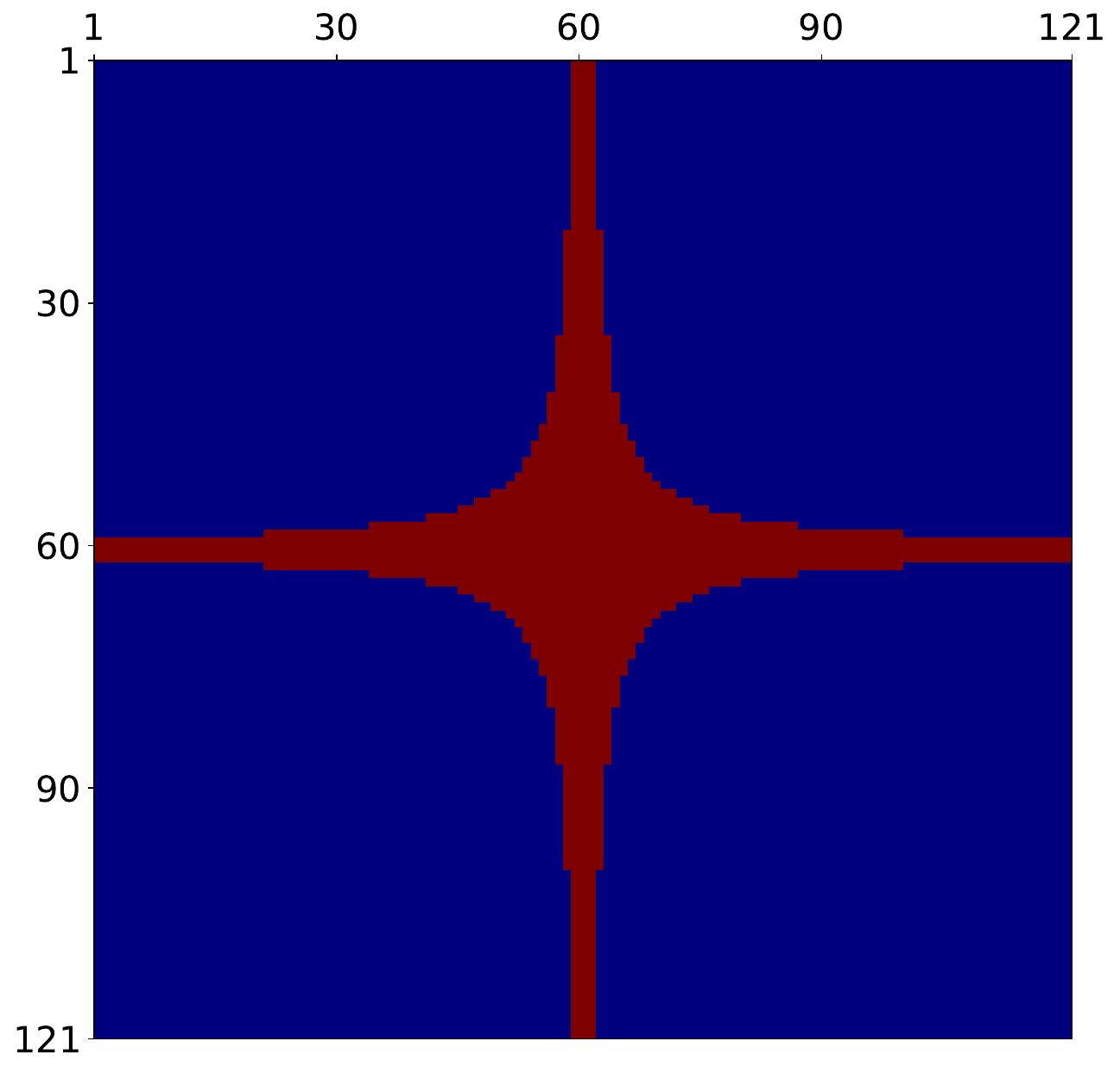}
		\caption{}
		\label{fig:attentionmaskblock1rho13t64}
	\end{subfigure}
	\begin{subfigure}{0.24\linewidth}
		\centering
		\includegraphics[width=.8\linewidth]{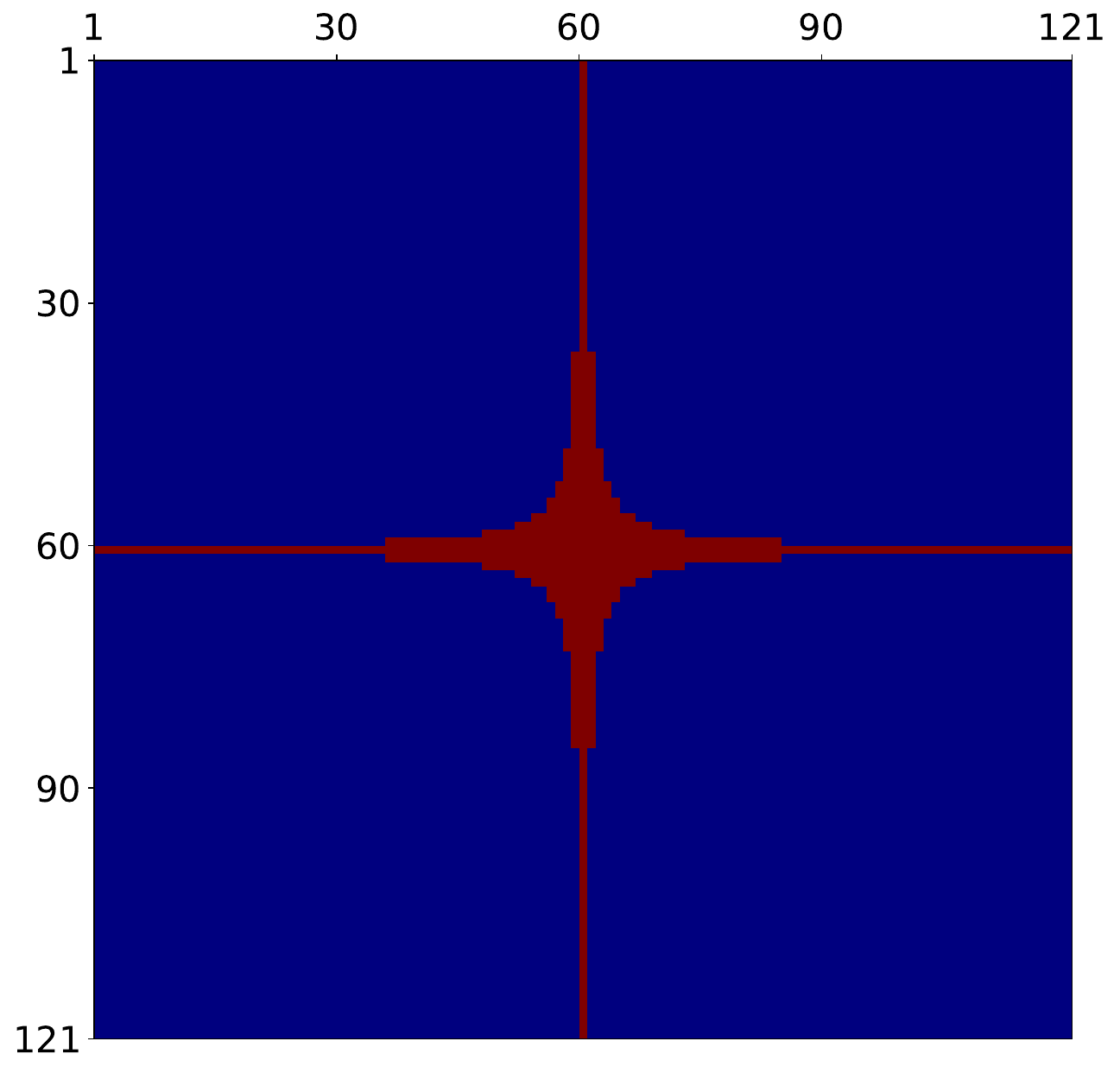}
		\caption{}
		\label{fig:attentionmaskblock1rho04t64}
	\end{subfigure}
	\caption{Attention masks for the cases where output error only computed for the individual target symbol. The red area shows the values that are zeros (unmasked) and the blue area shows the values which are negative infinity (masked). (a) $ t=32 $ and $ \rho=2.6 $, (b) $ t=64 $ and $ \rho=2.6 $, (c) $ t=64 $ and $ \rho=1.3 $, and (d) $ t=64 $ and $ \rho=0.4 $. The ratio of number of unmasked (zero) elements to the total number of elements is 0.31, 0.19, 0.10 and 0.04 for figure (a), (b), (c), and (d), respectively.}
	\label{fig:attentionmaskblock1}
\end{figure*}

At first, consider the mask generation for an individual target symbol (block size of one). The algorithm for generating $M^{indv}$ is shown in Alg.~\eqref{alg:mask_generation}.
Depending on the values of tap size $ t $ and mask hyperbolic parameter $ \rho $, the number of non-zero elements in the mask varies and consequently the number of operations needed for computing the attention matrix changes. 
Figure \ref{fig:attentionmaskblock1} shows the attention mask for the individual symbol processing for different values of $ t $ and $ \rho $.

\begin{algorithm}[b]
	\small
	\caption{\small Mask generation for an individual target symbol.}
	\label{alg:mask_generation}
	\begin{algorithmic}
		\Require $ \ell, \rho $
		\Ensure $M^{indv}$
		\State $M^{indv}$ = matrix of size $(\ell+1)\times(\ell+1)$
		\State $M^{indv}_{i, j}$ = $ -\infty $,  $ i, j \in 1 \cdots \ell+1$
		\For{$m$ = $-\ell/2$ to $\ell/2$}
		\For{$n$ = $-\ell/2$ to $\ell/2$}
		\If{$ m \ne 0 $ \textbf{and} $ |n| \le \min \left( \frac{\rho \lceil \ell/2 \rceil}{|m|}, \left \lceil \frac{\ell}{2} \right \rceil \right) $}
		\State$  M^{indv}_{\ell/2+m+1, \ell/2+n+1} = 0 $
		\EndIf
		\EndFor
		\EndFor
	\end{algorithmic}
\end{algorithm}
%

\begin{figure*}[]
	\centering
	\begin{subfigure}{0.22\linewidth}
		\centering
		\includegraphics[width=0.85\linewidth]{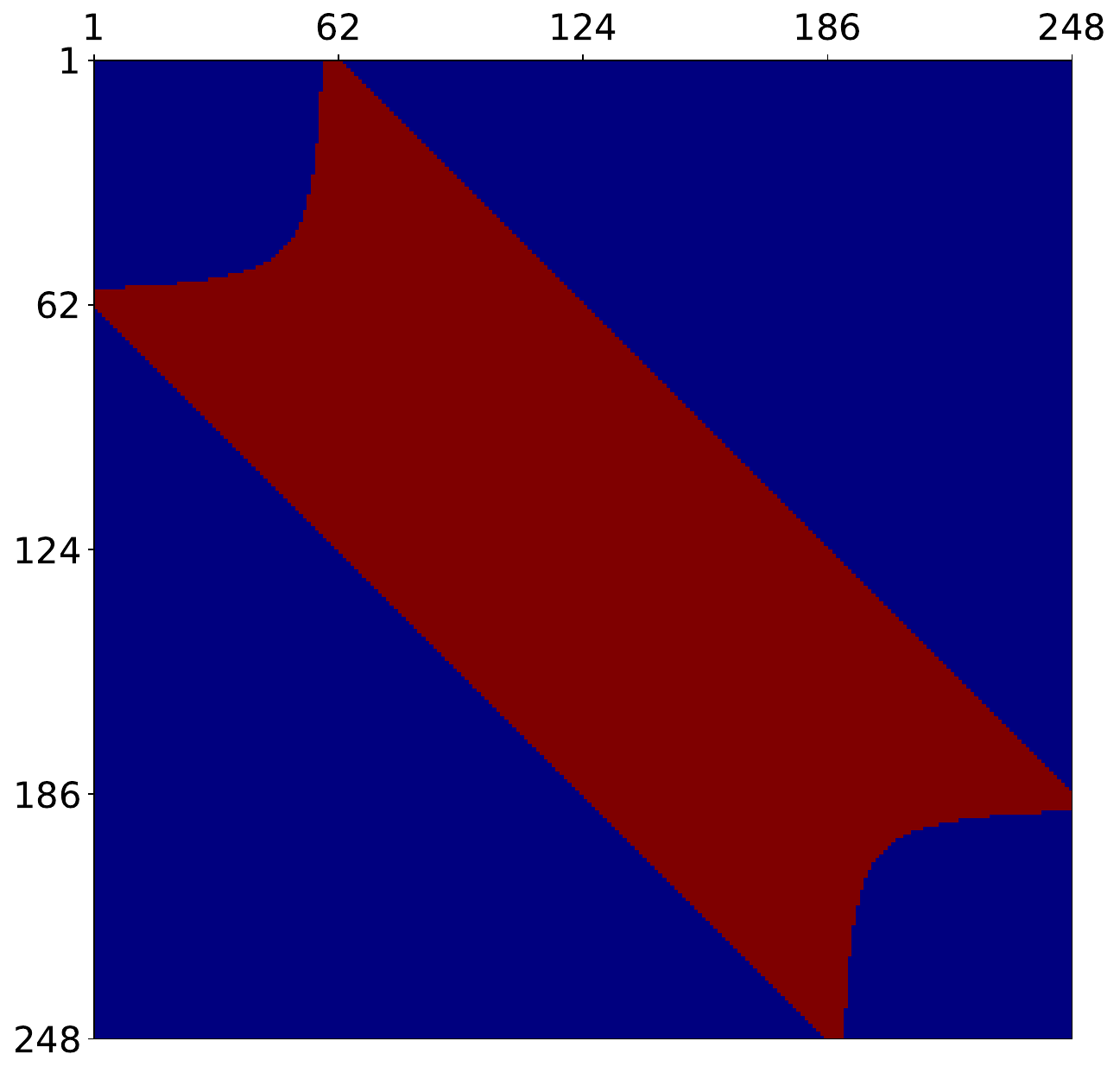}
		\caption{}
		\label{fig:attentionmaskblock128rho2p6tap64}
	\end{subfigure}
	\begin{subfigure}{0.22\linewidth}
		\centering
		\includegraphics[width=0.85\linewidth]{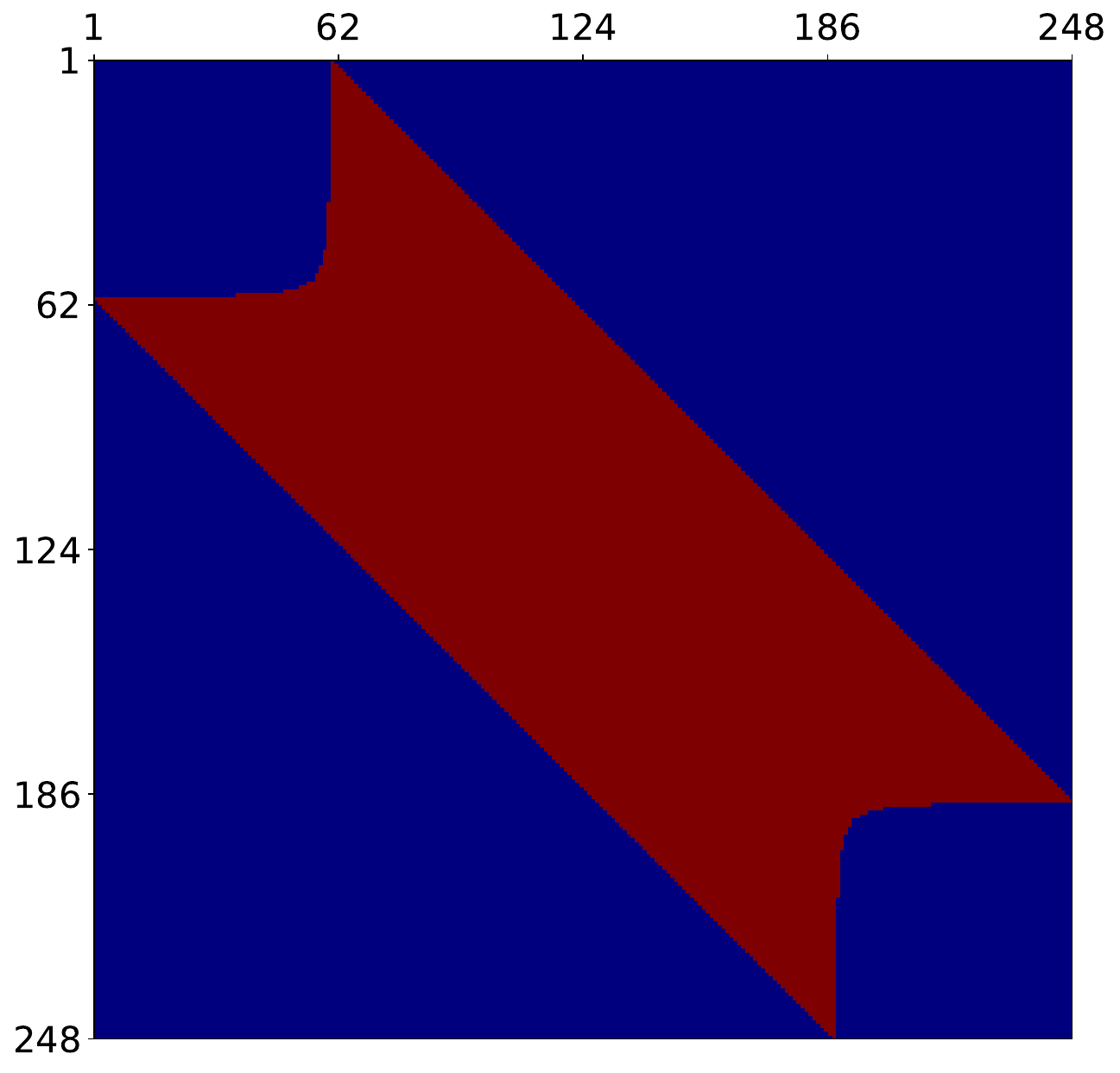}
		\caption{}
		\label{fig:attentionmaskblock128rho0p4tap64}
	\end{subfigure}
	\begin{subfigure}{0.22\linewidth}
		\centering
		\includegraphics[width=0.87\linewidth]{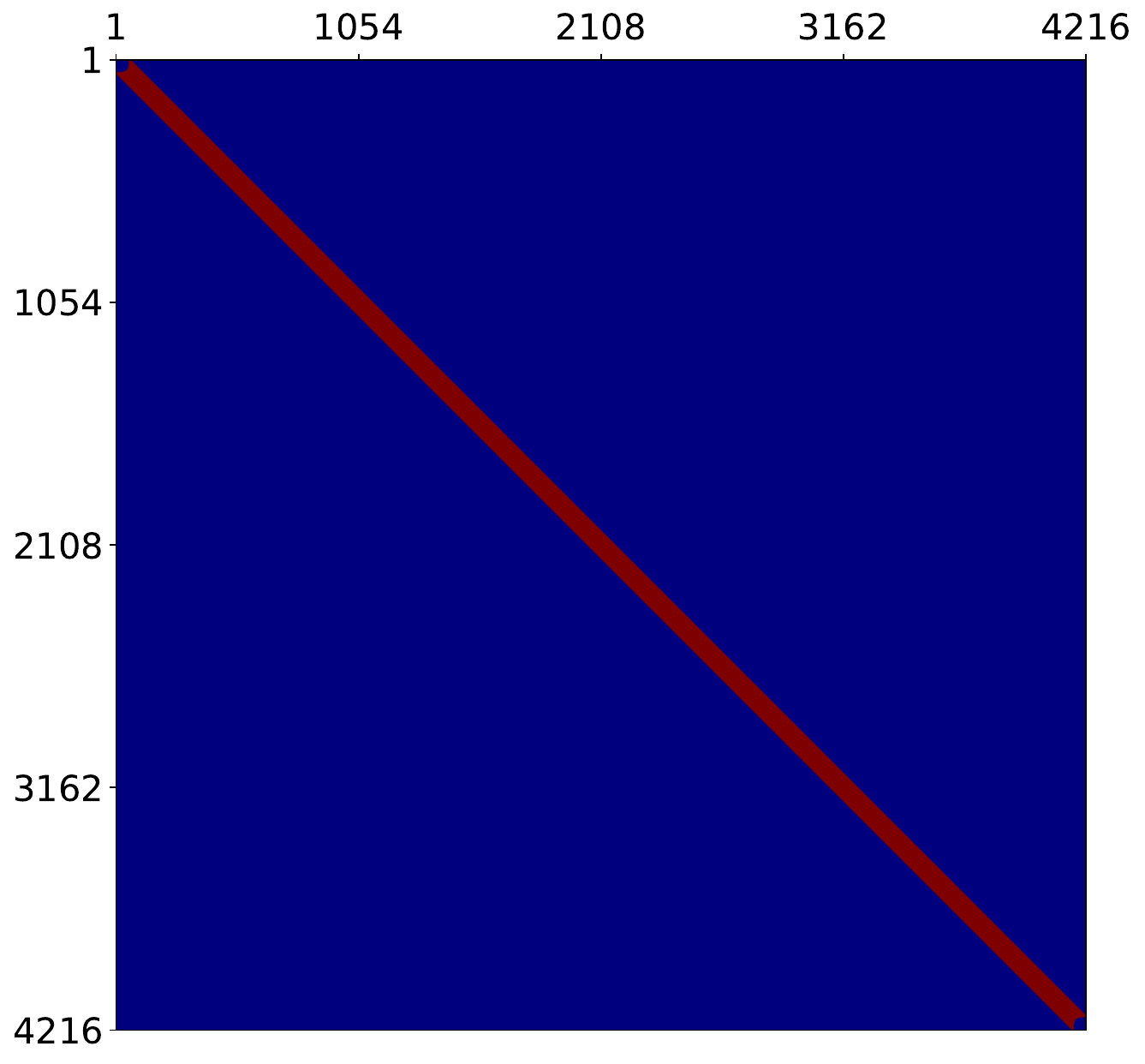}
		\caption{}
		\label{fig:attentionmaskblock4096rho2p6tap64}
	\end{subfigure}
	\begin{subfigure}{0.3\linewidth}
		\centering
		\includegraphics[width=.85\linewidth]{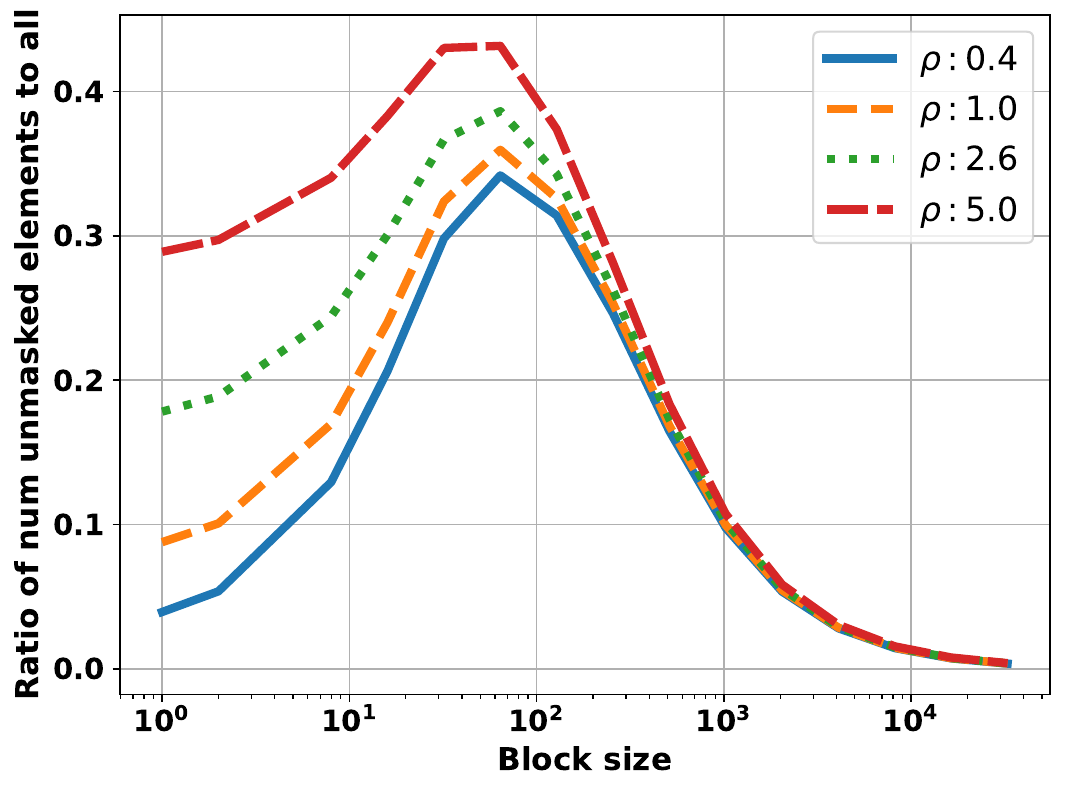}
		\caption{}
		\label{fig:zero_ratio}
	\end{subfigure}
	\caption{Masks for blocks: (a) shows a block mask with $t=64, b=128, \rho=2.6$ while (b) depicts a block mask with $t=64, b=128, \rho=0.4$. (c) shows the mask with $t=64, b=4096, \rho=2.6$. 
		The blue area shows the elements with negative infinity values and the red shows zero elements.
		(d) shows the ratio of number of unmasked (zero) elements to the total number of elements versus block size for several values of $\rho$s.}
	\label{fig:attentionmaskblock}
\end{figure*}

Next, we discuss the mask approach for the processing of a block of target symbols. 
The method we propose here is based on computing a union-like combination of the individual symbol's mask. The algorithm for computing the mask for the block-processing approach is shown in Alg.~\eqref{alg:block_mask_generation}.
Note that the size of the block mask matrix is $M^{block}$, is $(\ell+b) \times (\ell+b)$ while the size for individual symbol's mask was $M^{indv}$, is $(\ell+1) \times (\ell+1)$.
%
For two different values of $\rho$, block masks are depicted in Fig.~\ref{fig:attentionmaskblock} for a model with $t=64, b=128, \rho=2.6$ in Fig.~\ref{fig:attentionmaskblock128rho2p6tap64} and another model with $t=64, b=128, \rho=0.4$ in Fig.~\ref{fig:attentionmaskblock128rho0p4tap64}, respectively. 
As can be seen from this figure, the choice of $\rho$ does not significantly alter the mask which is not surprising since the block mask is generated by stacking individual masks on each others. 
The ratios of unmasked (zero) elements to the total elements are $0.34$ and $0.31$ for $\rho=2.6$ and $\rho=0.4$, respectively. 

\begin{algorithm}[b]
	\small
	\caption{\small Mask generation for a block of target symbols.}
	\label{alg:block_mask_generation}
	\begin{algorithmic}
		\Require $ \ell, \rho , b$
		\Ensure $M^{block}$
		\State $M^{block}$ = matrix of size $(\ell+b) \times (\ell+b)$
		\State $M^{block}_{i, j}$ = $ -\infty $,  $ i, j \in 1 \cdots \ell+b$
		\State $M^{indv}$ = generate mask for one symbol using Algorithm \ref{alg:mask_generation}
		\For{i = 1 to b}
		\State \begin{varwidth}[t]{\linewidth} 
			$M^{block}_{i:i+\ell, i:i+\ell}$ = $\underset{\text{\emph{element-wise}}}{max}(M^{block}_{i:i+\ell, i:i+\ell},  M^{indv})$
		\end{varwidth}
		\EndFor
	\end{algorithmic}
\end{algorithm}
%
%

For large block sizes, this ratio substantially decreases. 
Fig.~\ref{fig:attentionmaskblock4096rho2p6tap64} shows the impact of increasing the block size to 4096. In this case, only $3\%$ of elements of the attention matrix are preserved which results in a large complexity reduction. 
Finally, Fig.~\ref{fig:zero_ratio} shows the ratio of unmasked (zero) elements to the total number of elements as a function of block size for several values of $\rho$s. As the figure shows when the block size is increased the value of $\rho$ has less effects on the ratio of the number of unmasked elements to the total number of elements
In fact, the ratio first increases and then decreases as the block size increases. 
It should be noted that the choice of block-size needs to be optimized according to the system requirements in terms of throughput and latency, and hence, it cannot be increased too much due to increasing required memory and complexity of Transformers.
A detailed discussion on the impact of block size is given in \cite{hamgini2023application}. 
To give some numerical examples of the effect of mask on the softmax input, several two-dimensional heat maps of  $(QK^T)/\sqrt{d_k}$ for Transformer with and without masks are shown in Appendix \ref{sec:appendix_heatmap}.

\subsection{Using Neighbors at the Output Layer of Transformer}\label{Neighbor_output}
For computing nonlinear distortion from the output of the last layer of the Transformer, we employ an MLP as discussed in Section \ref{TF_overview}. 
%
To compute each nonlinear distortion $s_i$, one option is to give the corresponding Transformer-generated representation, $r_i$, to the MLP. However, due to the dependency of nonlinear estimates, we observe that if for each symbol $s_i$, we also provide few extra representations around $r_i$ (the \emph{neighbors}) to the MLP, it can compute $s_i$ more accurately.
We call this approach \emph{window processing} and while it can increase the model's performance, it does not increase the computational complexity substantially for small and moderate window sizes.
Specifically, the selected window of representations are fed into an MLP module consists of a linear layer ($ \in R^{W\cdot d_{model} \times 2 }$), a leaky ReLU nonlinear activation function with negative slope of $ 0.2 $, a second linear layer ($\in R^{2 \times 10} $), a leaky ReLU nonlinear activation function with negative slope of $ 0.2 $, and a third linear layer ($\in R^{10 \times 2 }$). The last layer has two outputs that generates the estimated nonlinear distortions corresponding to the target symbols.

Without window processing, the first layer of the MLP has $d_{model}$ inputs and hence, the computational complexity of the MLP is of order of $O(d_{model})$ per symbol.
With window processing, for each symbol $s_i$, in addition to the $r_i$, $w$ positions on each side of $r_i$ is also fed into the output MLP which will increase the complexity by order of $2w$.
However, since $w$ is fixed in our design, the computational complexity is still of order of $O(d_{model})$ per symbol. 
Finally, note that the use of neighboring representations at the output of Transformer can also be exploited by other methods such as using a one-dimensional CNN as the first layer of the output module.

\section{Numerical Results}\label{TF_sim}

\subsection{System Model}
\begin{figure*}[]
	\centering	
	\includegraphics[width=.9\linewidth]{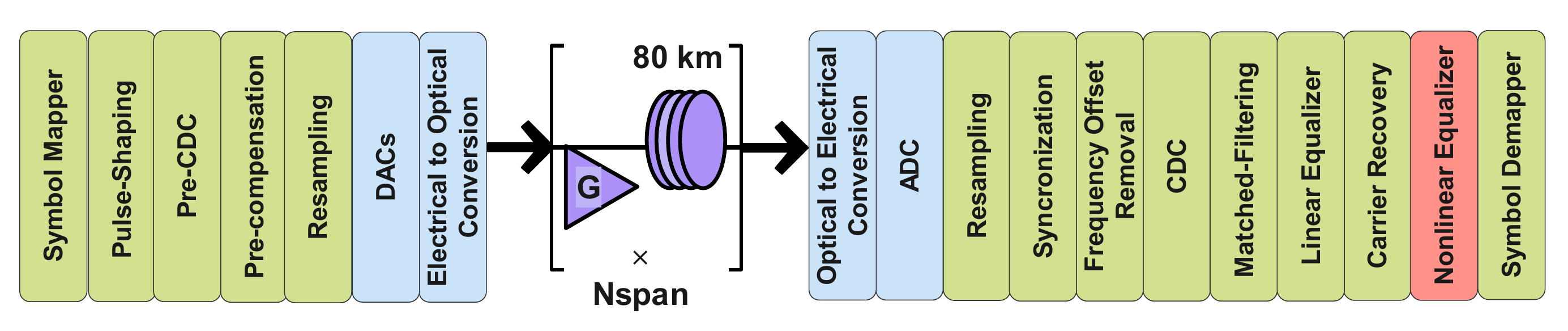}
	\caption{System model with nonlinear equalizer at the receiver side.}
	\label{diag:SC_system_baseline}
\end{figure*}

The system model for generating the simulation results includes typical Tx, channel, and Rx modules for single-carrier transmission. 
Specifically, we consider digital chromatic dispersion compensation modules both at Tx and Rx side. We also consider ideal electrical components and Mach-Zehnder modulator. Furthermore, DACs/ADCs are ideal with no quantization or clipping effects. 
The dual polarization fiber channel is modeled by split-step Fourier method \cite{SSFM} with adaptive step-size and maximum nonlinear phase-rotation of 0.05 degree to ensure sufficient accuracy. 
At the receiver side, standard DSP algorithms are employed for signal processing. The sequence output from carrier recovery (CR) are employed for training and evaluating Transformer-based nonlinear equalizer. 
Specifically, to maintain the capability of conventional coherent receiver for phase correction under correlated phase-noise we used CR before the ANN-NLC module.
In this manner, the linear processing has the nonlinear phase compensation ability of a coherent receiver without a dedicated NLC equalizer. 
Hence, the neural network compensation gain is given on top of the best linear performance. 
The block diagram for such system is depicted in Fig.~\ref{diag:SC_system_baseline}.

To study the performance versus complexity trade-off, we focus on two cases with single-channel coherent system. 
The first one is operating at 32 Gbaud 16QAM modulation where the link consists of 40 spans of standard single-mode fiber (SSMF).
The second case uses 42 Gbaud 64QAM modulation format with 12 spans of SSMF. Each span has a length of 80km followed by an optical amplifier with $6$ dB noise figure.
The signals are pulse shaped by a root-raised cosine function with a roll-off factor of $1/16$. 
Furthermore, we consider a symmetric dispersion map, in which half of total dispersion is digitally pre-compensated at the transmitter side.
Training is performed at $2$ dBm and $3$ dBm launch powers for the two cases, respectively. 
This is close to the optimal launch power for each case.

\subsection{Numerical Setup}
Standard Rx-DSP output for $2^{19}$ and $2^{18}$ symbols were used in the training and evaluation of ANN models, respectively. 
A permuted congruential generator (PCG64) with different seeds for training and evaluation stages was employed.
Models are trained for a mini-batch size of 512 using Adam optimizer with a learning rate scheduler that includes a warm-up stage, and a mean-squared-error loss function. Early stopping is used to stop the training if the model performance has not improved for over 100 epochs.
Note that the proposed ANN-NLC equalizers estimate the nonlinear distortions in one polarization. However, due to the symmetric nature of signal propagation in fiber medium, it has been shown that the same model can be used to generate the nonlinear error estimates for the other polarization by simply swapping input signals. This enables efficient learning of a generalized model that performs on both polarizations. Furthermore, the output layer can be slightly modified in order to generate the nonlinear estimates for both polarizations at the same time.

Parameters and components in a Transformer architecture can be varied in order to explore performance and computational complexity of different realizations. We perform a wide grid search over hyper-parameters of the Transformer and various components consisting of hundreds of models.
Among these parameters, we studied the impact of following hyper-parameters: tap size, $d_{model}$ which we will call \emph{hidden size}, key size, number of heads, number of encoder layers, number of outputs for the first layer of point-wise FFN which is called \emph{FFN's hidden size}, presence or absence of mask, and output layer window size. 
We implemented the search in two steps: 
At first, for each hyper-parameter we ran a specific grid search where we selected a wide range for that specific hyper-parameter while only few values for the remaining parameters were included. In the second step, we chose some of the best values from the first run and performed a grid search on the combination of those values. 
As discussed earlier, the type of network that generates the embeddings is important. Hence, we carried out a separate grid-search on the impact of block size and embedding generator networks on a smaller grid where the details are presented in \cite{hamgini2023application}. 
%
Table \ref{tbl:grid_searchqam} shows the summary of selected values for the grid search in the second step. 
We have employed separate search spaces for 16QAM and 64QAM cases tailored to their propagation mediums.

\begin{table}[t]
	\centering
	\caption{Transformer's hyper-parameters used for the grid search.}
	\label{tbl:grid_searchqam}
	\footnotesize
	\renewcommand{\arraystretch}{.95}
	\begin{tabular}{p{2.8cm}p{2.2cm}p{2.2cm}}
		\Xhline{1pt}
		\vspace{1pt}
		\textbf{Hyper-parameter}& 
		\vspace{1pt}
		\textbf{Grid-Search \mbox{Values} (16QAM)}  & 
		\vspace{1pt}
		\textbf{Grid-Search \mbox{Values} (64QAM)}\\
		\hline
		tap size &  $\in$ [8:96] &  $\in$ [16:64] \\
		hidden size & $\in$ [8:96] &  $\in$ [16:96] \\
		key size &  $\in$ [8:64] & $\in$ [16:64]\\
		number of heads & 1, 2, 4& 1, 4\\
		number of encoder layer & 1, 2, 3 & 1, 2, 3\\
		FFN's hidden size & 32, 64& 64\\
		window size & 1, 7, 15 & 1, 7, 11\\
		mask & None , $\rho=2.6$ & None , $\rho=2.6$\\
		\Xhline{1pt} 
	\end{tabular}
\end{table}

\subsection{Performance vs. Complexity Trade-off for 16QAM Setup} \label{Numerical_TF_16qam}

\begin{figure}[t]
	\centering
		\centering
		\includegraphics[width=.65\linewidth]
		{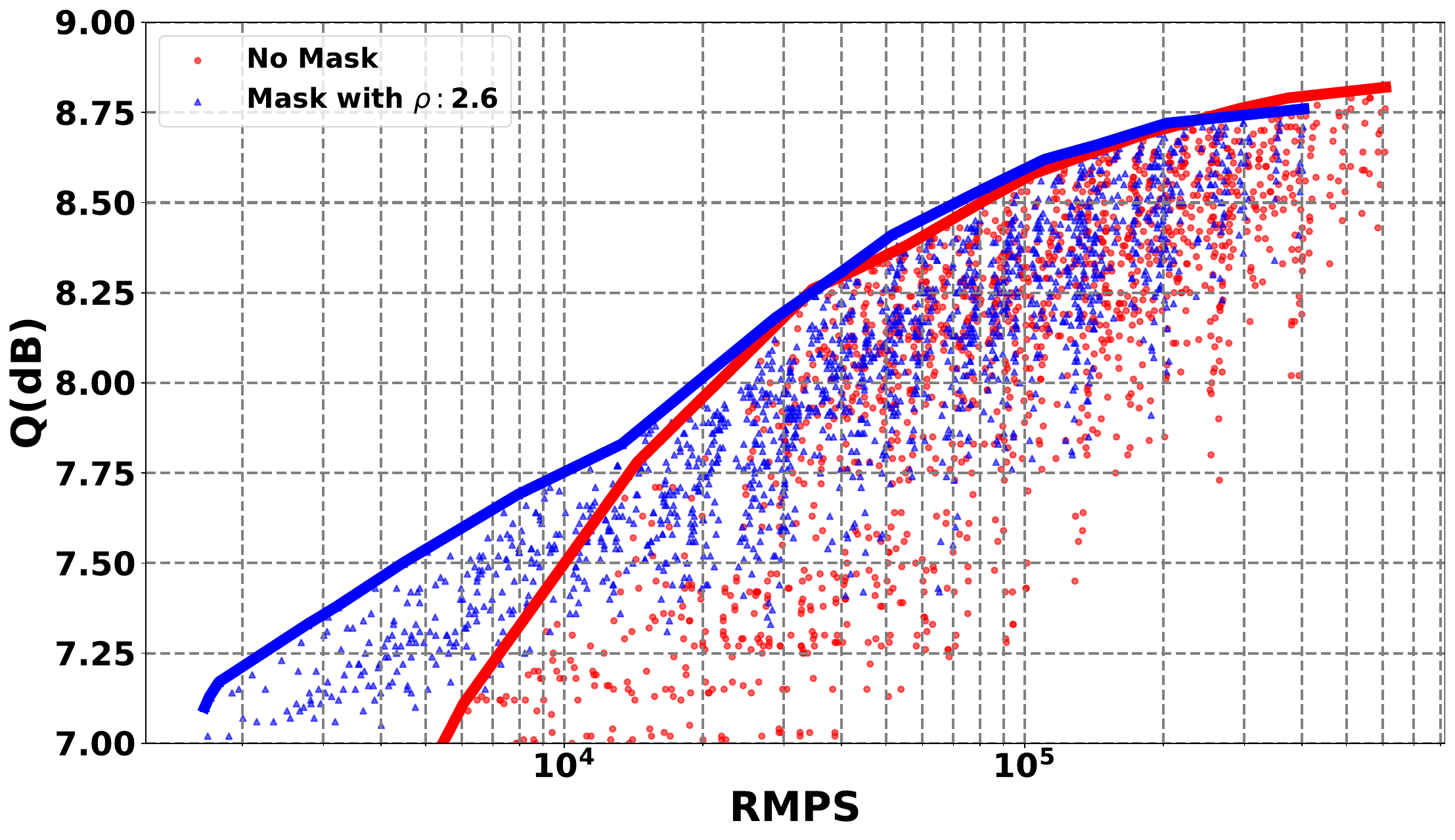}
		\caption{Performance vs. complexity trade-off for the 16QAM setup with grid search shown according to Table \ref{tbl:grid_searchqam}. The models were trained and evaluated at $2$ dBm launch power. 
			Red dots show the results for models that did not use any mask while blue triangles represent the ones with the proposed mask.
		}
		\label{fig:tfrhoallQvsFlopenvelopesweepscattertogether}
\end{figure}

The numerical results for hyper-parameters grid search in the 16QAM case is given in Fig.~\ref{fig:tfrhoallQvsFlopenvelopesweepscattertogether}. 
This figure shows the scatter plot for the performance, represented by the quality metric Q, versus complexity, represented by RMPS, for all model variations obtained from sweeping over hyper-parameters in Table~\ref{tbl:grid_searchqam}. We use blue and red colored dots to show the results for Transformers with and without mask, respectively. Additionally, the blue and red curves illustrate the envelopes associated to the best performing Transformers with and without mask, respectively. 
All Transformer-NLCs are designed for a block size of 128 symbols.
As seen from the figure, Transformers can achieve impressive nonlinear compensation performance as the best models in 16QAM setup have improved Q from $6.7$ dB (linear case) to more than $8.8$ dB, which is $2.1$ dB improvement at the training launch power of $2$ dBm. 
In addition, even at the lower complexities where the performance has been decreased, they still retain a good nonlinear compensation capability. 
Moreover, using a mask particularly boost the performance at lower complexity regions compared to the case without mask while there is no degrading impact on the performance at higher complexities.

It is worthwhile to note that at mid and high complexity regions, the proposed mask helps the training process by pointing to the most important attention positions and hence, it can save in computational complexity for a given well trained model. 
However, for the lower complexity region, there is a performance advantage as well.
In fact, directing limited resources toward important model parameters provides a learning advantage here where one can see the performance boost in this region. 
In other words, with under-parameterized models in low complexity region, the mask guidance generally results in a more efficient training process.  Table \ref{tbl:16QAMThreeSampleModels} shows the performance and complexity of three models as examples of different performance versus complexity regions from Fig. \ref{fig:tfrhoallQvsFlopenvelopesweepscattertogether}. 

\begin{table}[b] 
		\centering
		\caption{Impact of applying mask on performance and complexity of three selected models at 2 dBm launch power for 16QAM. }
		\footnotesize
		\begin{tabular}{llllcllc}
				\Xhline{1pt} 
				& & \multicolumn{2}{c}{\textbf{No Mask}} & \multicolumn{2}{c}{\textbf{Mask with $\bm{\rho=2.6}$}} \\ \cline{3-4} \cline{5-6} 
				\textbf{Model} & \text{Q(dB)} & Q\textsubscript{NN}(dB)    &  \text{RMPS(K)}   &  Q\textsubscript{NN}(dB)   & \text{RMPS(K)}  \\   \cline{1-2} \cline{3-4} \cline{5-6} 
				A & 6.67 &  8.82 & 608 &  8.76 & 404 \\ 
				B & 6.67 &  8.54 & 129 &  8.51 & 81 \\ 
				C & 6.67 &  7.31 & 22 &  7.68 & 9 \\  \Xhline{1pt} 
			\end{tabular}
		\label{tbl:16QAMThreeSampleModels}
\end{table}

Finally, note that a deeper study on impact of different hyper-parameters on performance versus complexity of the proposed Transformer-NLC has been presented in \cite{hamgini2023application}. The details are omitted here for brevity.

\subsection{Performance vs. Complexity Trade-off for 64QAM Setup}
\begin{figure}[t]
	\centering
	\includegraphics[width=.65\linewidth]{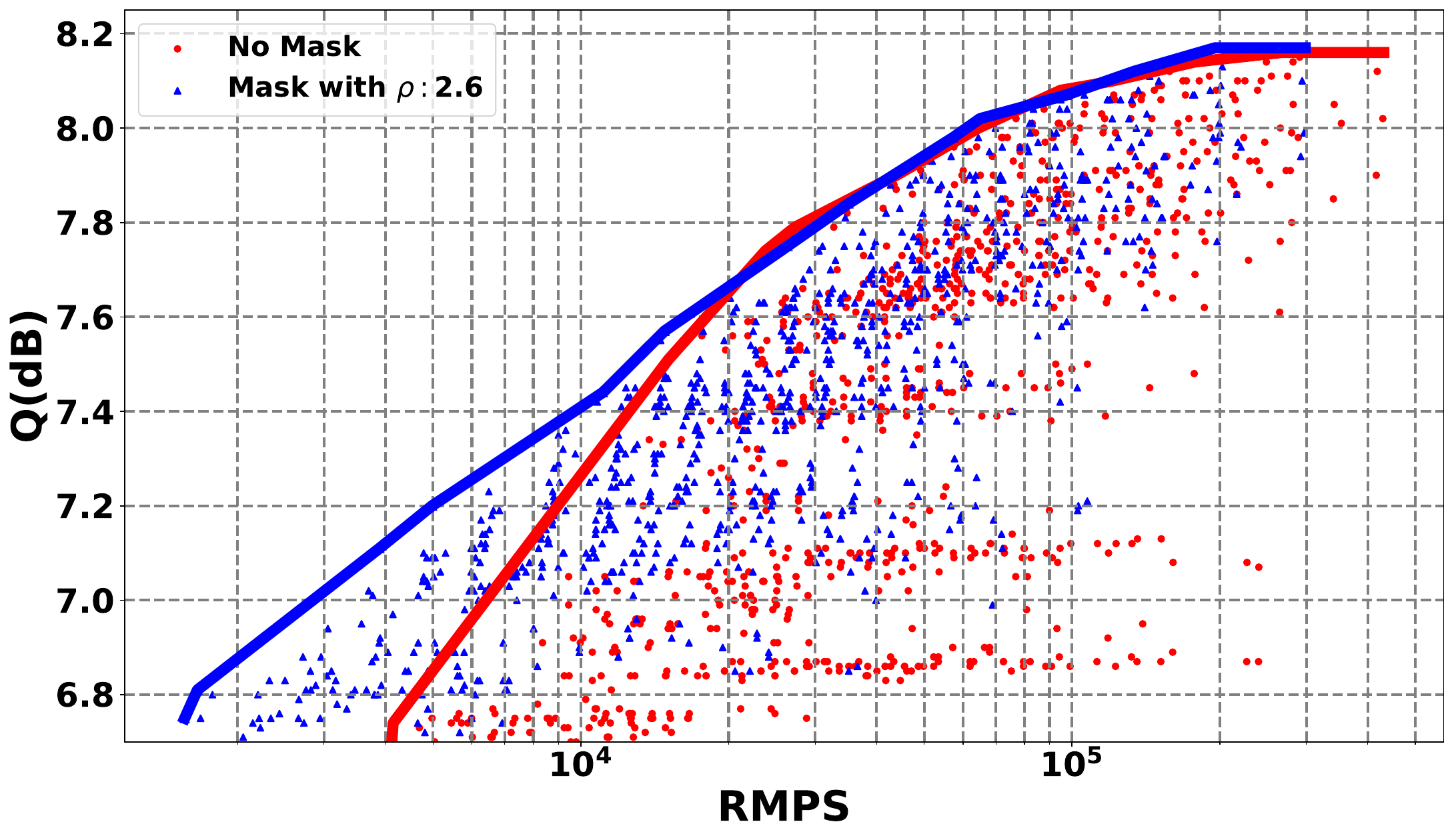}
	\caption{Performance vs. complexity trade-off graph of the 64QAM setup for the grid search shown in Table \ref{tbl:grid_searchqam}. Models were trained and evaluated at 3 dBm launch power. Red dots (blue triangles) represent models with (without) the proposed mask, respectively.}
	\label{fig:tfrhoallqvsflopconvexsweep64qam2setsscattertogether}
\end{figure}

As for the previous case, we also perform a grid search for 64QAM setup. 
All models are trained with a block size of 128 symbols. 
Fig.~\ref{fig:tfrhoallqvsflopconvexsweep64qam2setsscattertogether} shows the scatter plot for the performance versus complexity, for all model variations by sweeping over the hyper-parameters for 64QAM in Table~\ref{tbl:grid_searchqam}. 
In the figure, blue and red colored dots are used to show the results for Transformers with and without mask, while the blue and red lines are employed to depict the envelopes associated to the best performing Transformers with and without mask, respectively. 
We observe that the Transformers provide an excellent nonlinear compensation performance as the best models have improved Q from 6.5 dB (linear case) to 8.17 dB, a 1.67 dB improvement at the training launch power of 3 dBm. 
Also in this setup, the attention mask boosts the performance at lower complexity regions with almost no loss in general.

\subsection{Comparison to an LSTM-based equalizer} \label{sec:appendix_comparison}
As discussed in Section \ref{intro}, RNNs and specifically LSTMs are extensively studied for fiber nonlinearity equalization.
Combination of CNN and LSTM (CNN-LSTM) generally shows better performance compared to using LSTMs alone despite incurring additional complexity.
Also,  more efficient LSTM structures have been introduced for single carrier and digital subcarrier multiplexing structures to use block-processing for equalization stage. 
To do this, the models are trained on one symbol and executed on longer blocks of symbols which results in reduction of the computational complexity \cite{co-lstm, Bakhshali2023NeuralDMB}. 

\begin{figure}[b]
	\centering
	\includegraphics[width=0.65\linewidth]{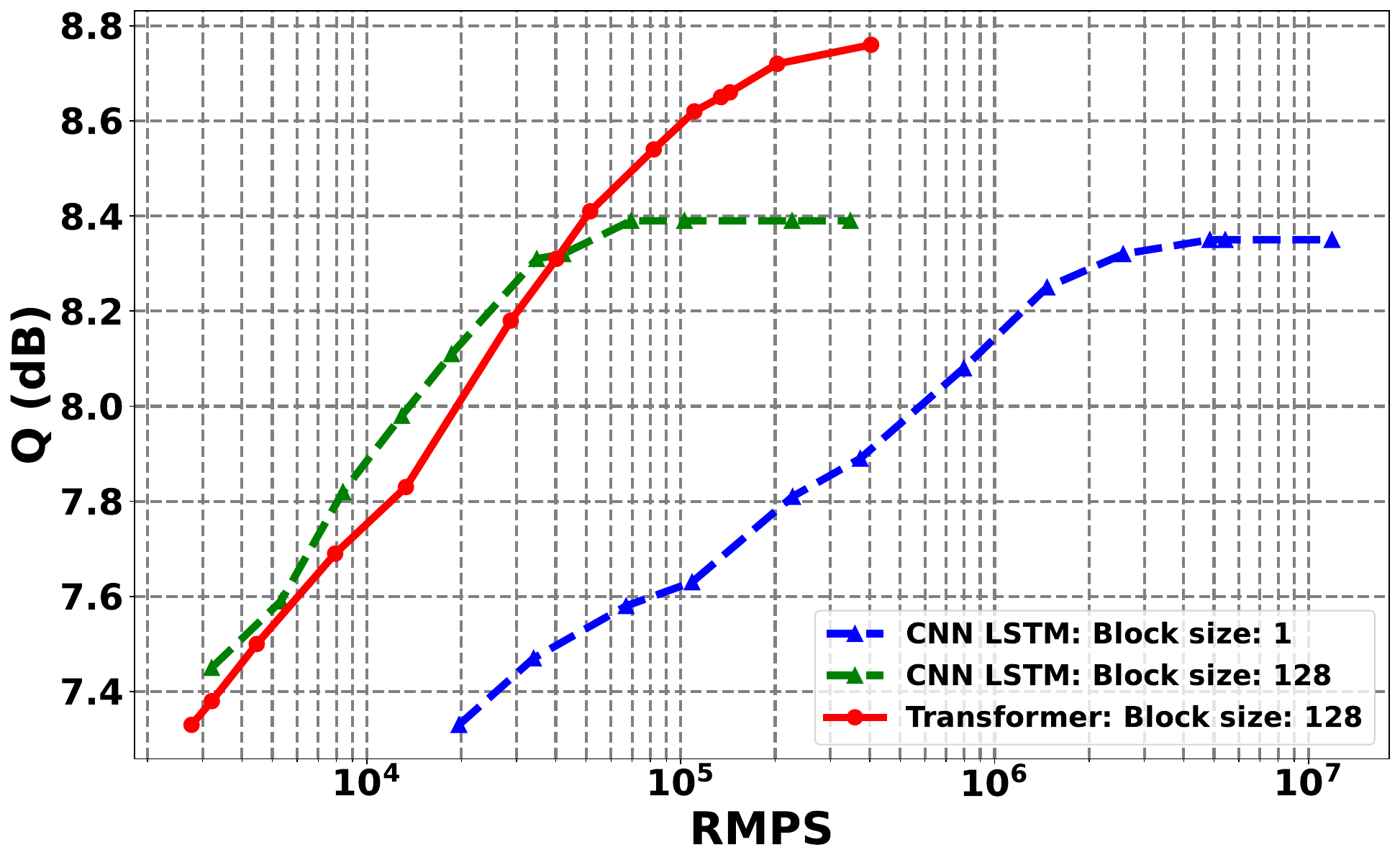}
	\caption{The envelope of performance vs. complexity for NLCs with Transformer and CNN-LSTM models. 
	}
	\label{fig:ncnncolstmvstransformerperfcomplex}
\end{figure}

Although a comprehensive comparison between the proposed Transformer model and CNN-LSTM structures is beyond the scope of this paper, we compare them through an example in order to demonstrate the capability of Transformers for our application. 
We use an optimized CNN-LSTM structure according to the details given in Section 3 of \cite{Bakhshali2023NeuralDMB}. 
Table \ref{tbl:grid_searchcnnlstm} summarizes the grid search regions for the hyper-parameters of the CNN-LSTM model.
%
%
We also use the same system model as shown in Fig. \ref{diag:SC_system_baseline} for both networks with the block size of 128. 
Note that while the Transformer models are trained and evaluated at the same block-size, CNN-LSTMs are trained at block-size of 1 and evaluated at block-size of 128. 
Fig.~\ref{fig:ncnncolstmvstransformerperfcomplex} shows envelopes of the evaluated models' performance versus complexity trade-off over different RMPS regions.
The result for CNN-LSTM models with block-size equal to one is also depicted here to show how block-processing in LSTM can improve performance despite incurring much higher latency.   
Considering just the float-point real multiplication and ignoring latency, one can see that at higher complexities, the Transformer outperforms CNN-LSTM by a large margin (more than 0.3 dB).
At lower complexities, CNN-LSTM models demonstrate a slight advantage, however, this amount is not consistent at all regions and the gap is small. 
We hypothesize that at higher complexities, the attention mechanism is more powerful and can capture the system memory better than the gated LSTM structure. However, at lower complexities, the attention's capacity is not large enough to capture the dependencies very well among the 128 symbols compared to the folding mechanism in LSTM.

\begin{table}[t]
	\centering
	\caption{CNN-LSTM's hyper-parameters and used for the grid search.}
	\label{tbl:grid_searchcnnlstm}
	\small
	\renewcommand{\arraystretch}{.95}
	\begin{tabular}{lc}
		\Xhline{1pt}
		\textbf{Hyper-parameter}& \textbf{Grid Search Values} \\
		\hline
		tap size &  $\in$ [10:90] \\
		hidden size & $\in$ [5:120] \\
		CNN's output filters &  $\in$ [10:90]\\
		MLP layers (middle MLP size) & 1 (0), 2 (32), 2(64)\\
		\Xhline{1pt} 
	\end{tabular}
\end{table}

In general, the complexity of LSTMs is roughly proportional to $(b+\ell)h^2/b$ where $(b+\ell)$ is the input sequence length, $h$ is the hidden size, and $b$ is the block size \cite{Bakhshali2023NeuralDMB}, while as expressed in this paper, the complexity of the attention is proportional to $(b+\ell)^2h/b $, which is quadratic in length of the input sequence. Therefore, as we increase the block size, the complexity eventually increases linearly in Transformers. 
Also, note that the employed dot-product attention may not be the best solution when it comes to lower complexity budgets. Several other architectures have been studied to reduce the complexity of the attention section from quadratic to linear. 
Incorporating these methods to reduce the complexity of Transformers can also be a research direction for future implementations.
Finally, as discussed before, the latency is not investigated here where the sequential structure of folding memory in LSTM introduces significant implementation challenges for coherent optical applications compared to the parallel attention mechanism in Transformers.

\begin{figure*}[t]
	\begin{subfigure}{0.32\linewidth}
		\includegraphics[width=1\linewidth]{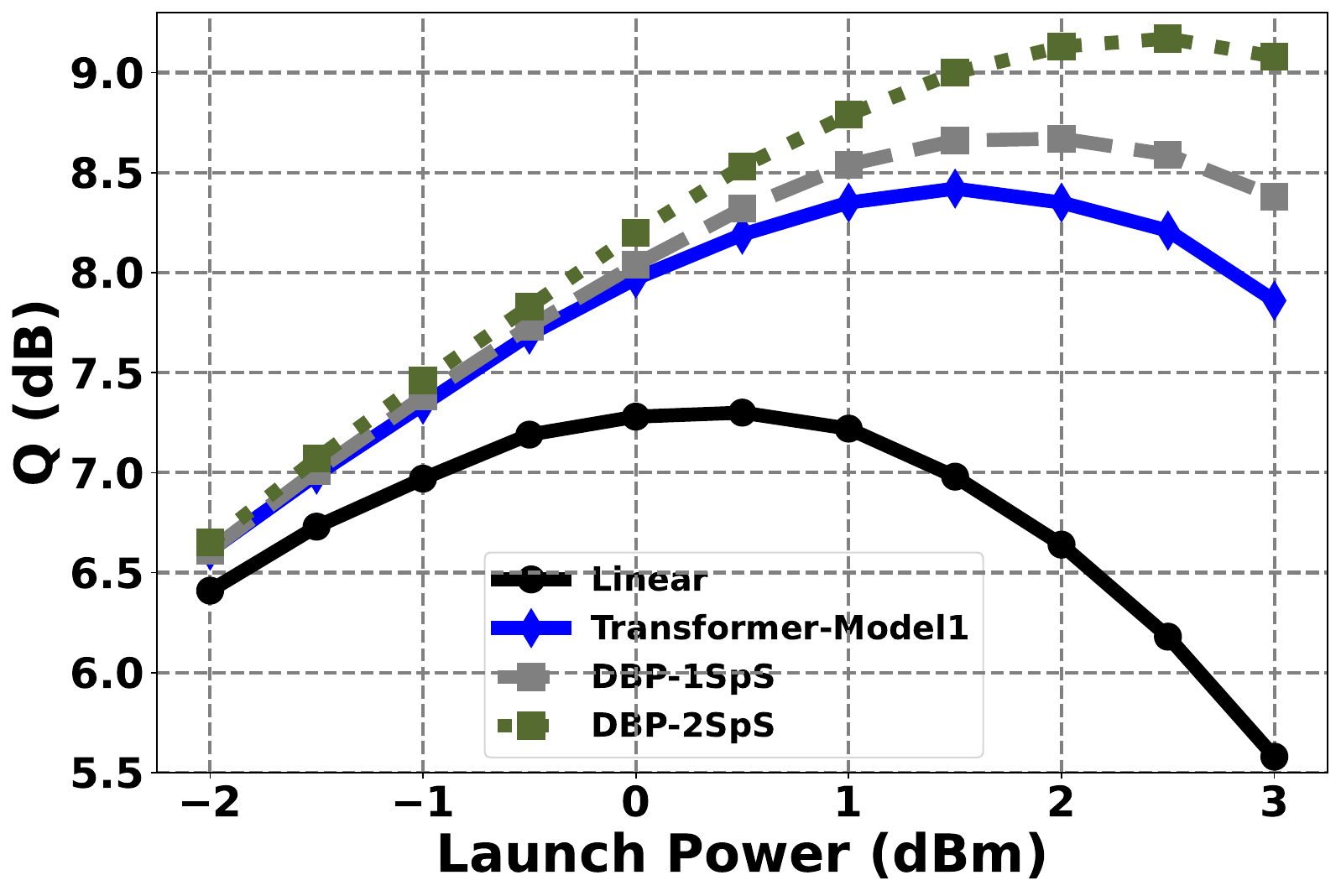}
		\caption{}
		\label{fig:powersweeep16qam32gb}
	\end{subfigure}
	\begin{subfigure}{0.32\linewidth}
		\includegraphics[width=1\linewidth]{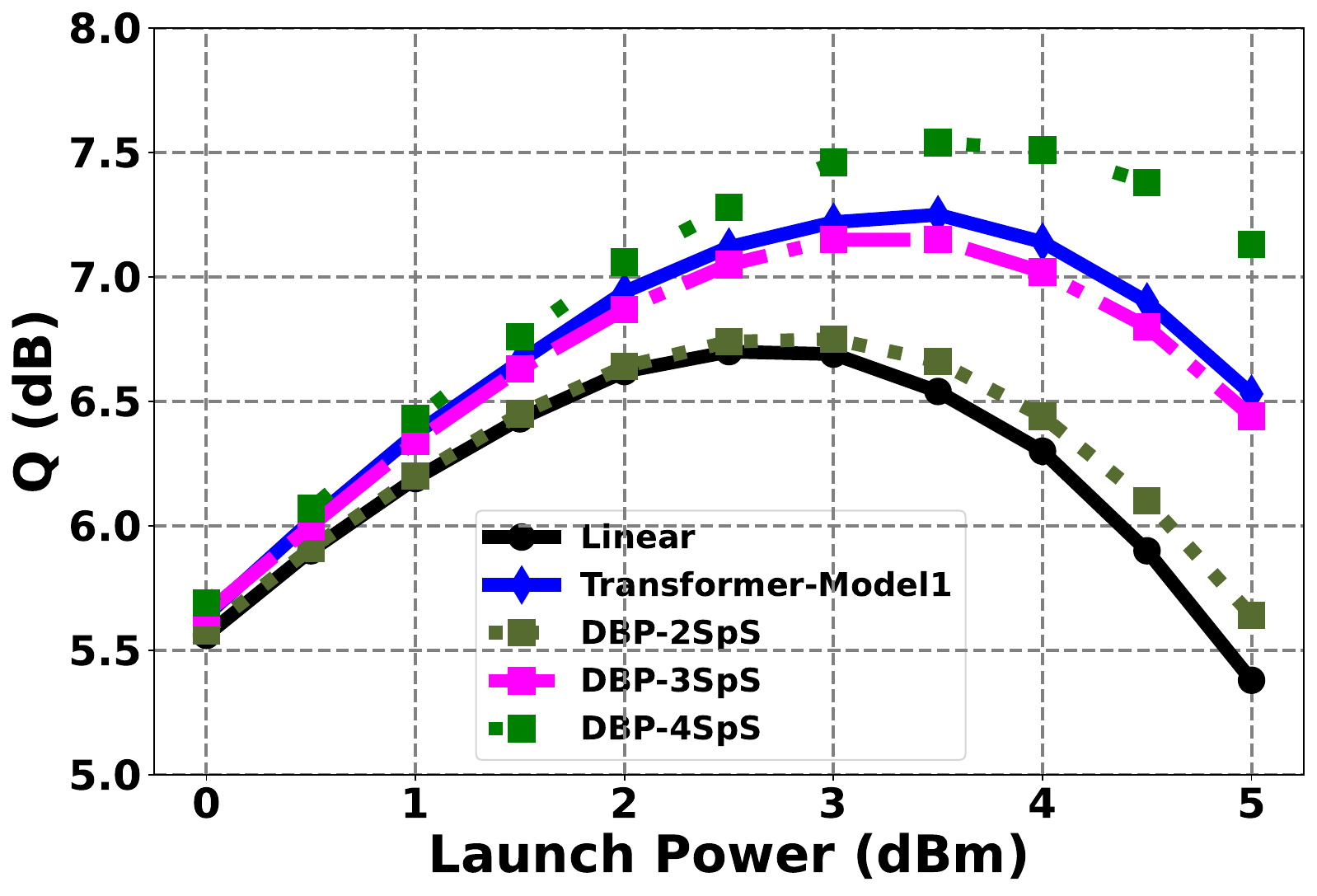}
		\caption{}
		\label{fig:powersweeep_16QAM_64GB}	
	\end{subfigure}
	\begin{subfigure}{0.32\linewidth}
		\includegraphics[width=1\linewidth]{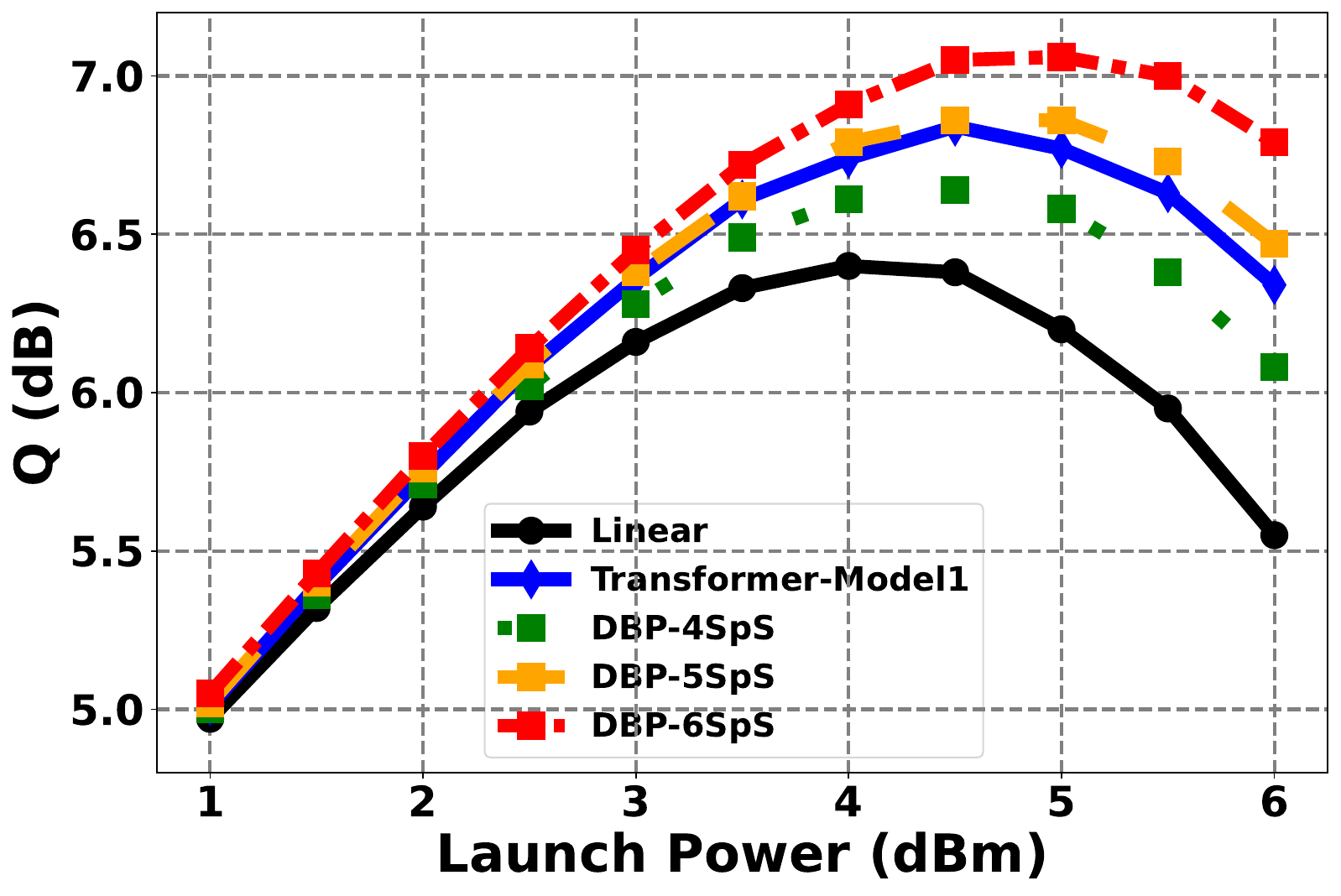}
		\caption{}
		\label{fig:powersweeep_16QAM_96GB}	
	\end{subfigure}
	\caption{Impact of symbol rate: performance of Model1 of Table \ref{tbl:selectedmodelshpyerarams} compared with DBP with various steps-per-span at several launch powers. Figs (a), (b), and (c) show the results at 32, 64, 96 GBaud, respectively.}
	\label{fig:compare_32_64_96GB}
\end{figure*}

\subsection{Comparison to DBP} \label{sec:comparison-to-dbp}
It is well-known that DBP faces serious challenges for implementation in high-speed coherent
transceivers including requiring higher over-sampling rate, lack of flexibility and generalization to the transmission scenario such as the need for more number of steps-per-spans with increasing symbol rate or at higher polarization mode dispersion (PMD) to maintain the performance. Also, use of multiple (inverse) fast Fourier transform (FFT) modules impacts the linear equalization data path.
Furthermore, as found in \cite{sherborne2018impact}, even the low-complexity DBP solutions are sensitive to quantization noise and require a high number of bit depth in order to improve over linear equalization. However, the robustness and effectiveness of neural networks under quantization and pruning are vastly shown in the literature as an advantage of such data-driven models \cite{QNN1,gholami2022survey,qin2020binary,zhu2017prune,hamgini2024pruning}. 

Although a complete comparison between DBP and Transformer-NLCs is beyond the scope of this paper, we use two transmission scenarios in this section as examples to show the generalization advantage of Transformer-NLCs.
We select two Transformer-NLC models close to the envelope of performance versus complexity regions of Fig. \ref{fig:tfrhoallQvsFlopenvelopesweepscattertogether} and fix the model parameters for all transmission scenarios
to compare them with DBP. 
The hyper-parameters for these models are shown in Table \ref{tbl:selectedmodelshpyerarams}.
Model1 is selected from the middle of performance versus complexity region while Model2 is chosen from the lower region. Both Transformer models use physic-informed masks with $\rho=2.6$.
Performance results for DBP are presented at two samples-per-symbol with different steps-per-span~(SpS) to benchmark the proposed Transformers-NLCs.  
Also, the transformer is trained in one launch power and is used in other launch powers by scaling according to \eqref{ANN_scale}.
Furthermore, without loss of generality, we modified the Transformer's output module in this section to generate the equalizer's nonlinear distortion estimations for both polarizations in order to save in the overall computational complexity of Transformer-NLCs.

\subsubsection{Impact of Increasing Symbol Rate}

In this part, we investigate the performance of Transformer-NLC and DBP at different symbol rates.
%
For this comparison, Model1 is considered for 32, 64, and 96 GBaud transmission scenarios. 
Note that this model is close to the envelope of performance versus complexity trade-off in the middle region for 32 GBaud transmission and may not be an optimized model for higher baud-rate scenarios. 

The performance results for the given Transformer-NLC model is shown in Fig.~\ref{fig:compare_32_64_96GB}.
The nonlinear compensation gains for the optimized DBP structures of various complexities operating at two samples-per-symbol are also depicted in these graphs.
We can observe that the nonlinear compensation is generally more challenging at higher baud rates as the gains are shrinking. 
Moreover, one need to increase the complexity of DBP and use more steps-per-span to match the given Transformer-NLC model.
This is in fact a critical issue with the application of DBP for high-speed optical transceivers (as in \cite{tan2019effectiveness} and references therein) and these results confirm the advantage and flexibility of a data-driven ANN model with increasing symbol rates over the deterministic method of DBP.

\begin{table}[]
	\caption{Hyper-parameters of the two selected models of 16QAM setup for launch power sweep.}
	\label{tbl:selectedmodelshpyerarams}
	\centering
	\small
	\renewcommand{\arraystretch}{1}
	\begin{tabular}{lcc}
		\Xhline{1pt}
		\textbf{Hyper-parameter} & \textbf{Model1}&\textbf{Model2} \\
		\hline
		tap size &32  &12 \\
		hidden size  &48  &16 \\
		key size  &16  &16 \\
		number of heads  &4  &4 \\
		number of encoder layer &3  &1 \\
		FFN's hidden size & 32  &32 \\
		window size &7  &7 \\
		\Xhline{1pt}
	\end{tabular}
\end{table}

\subsubsection{Impact of Increasing PMD}
\begin{figure*}[t]
	\begin{subfigure}{0.32\linewidth}
		\includegraphics[width=1\linewidth]{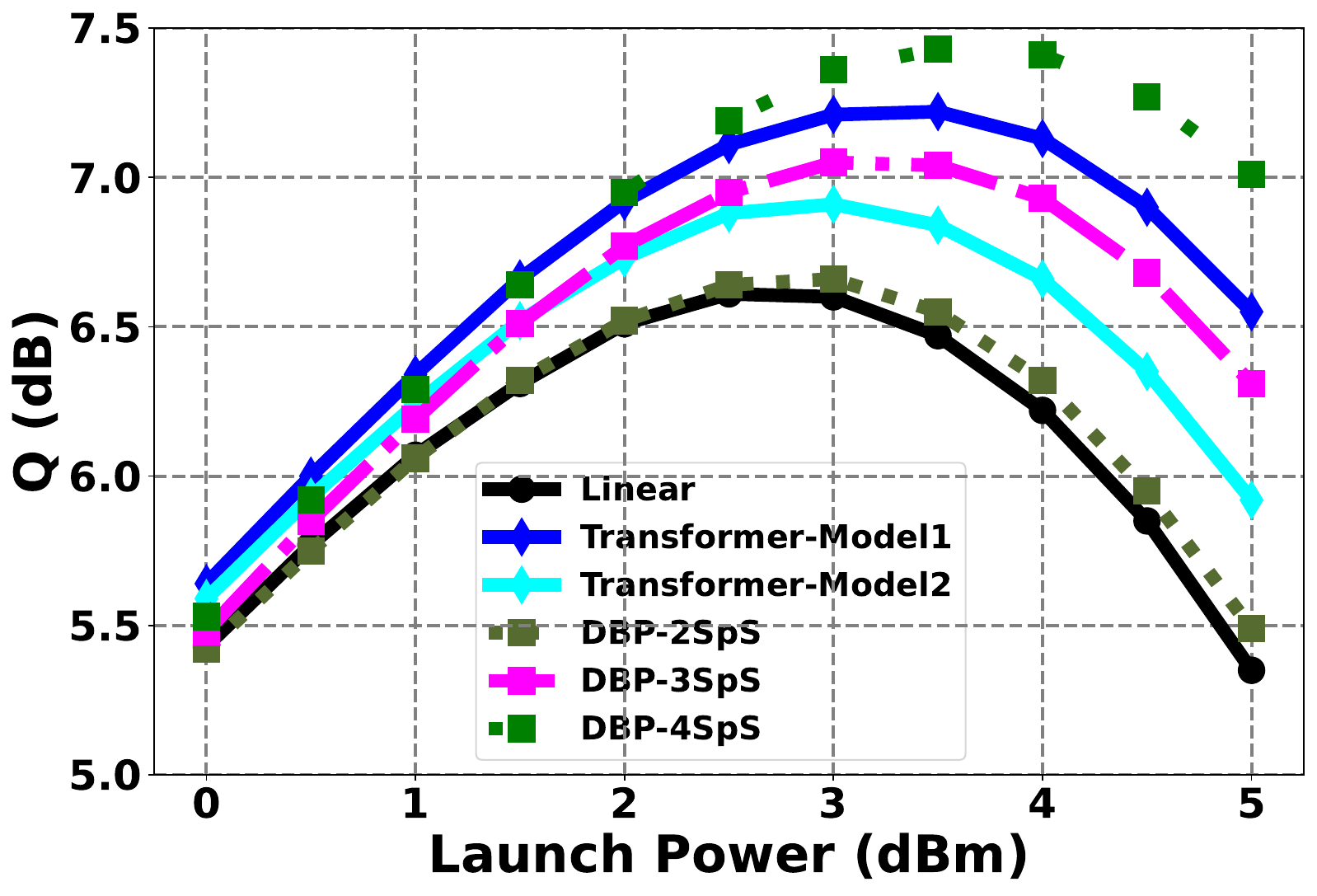}
		\caption{}
		\label{fig:transformerVsDbpPmd0d05K0d8Baud64GB}
	\end{subfigure}
	\begin{subfigure}{0.32\linewidth}
		\includegraphics[width=1\linewidth]{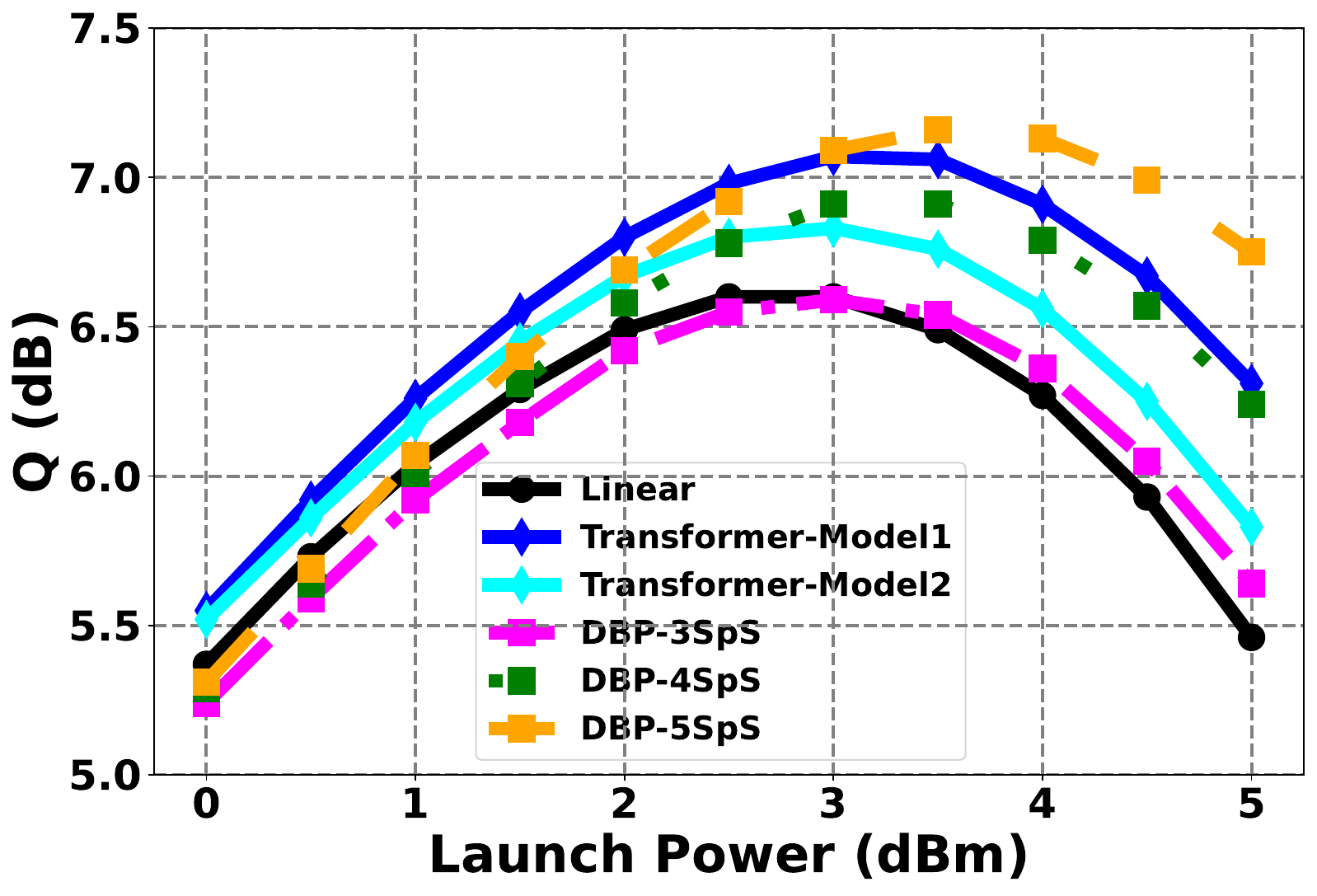}
		\caption{}
		\label{fig:transformerVsDbpPmd0d5K0d8Baud64GB}	
	\end{subfigure}
	\begin{subfigure}{0.32\linewidth}
		\includegraphics[width=1\linewidth]{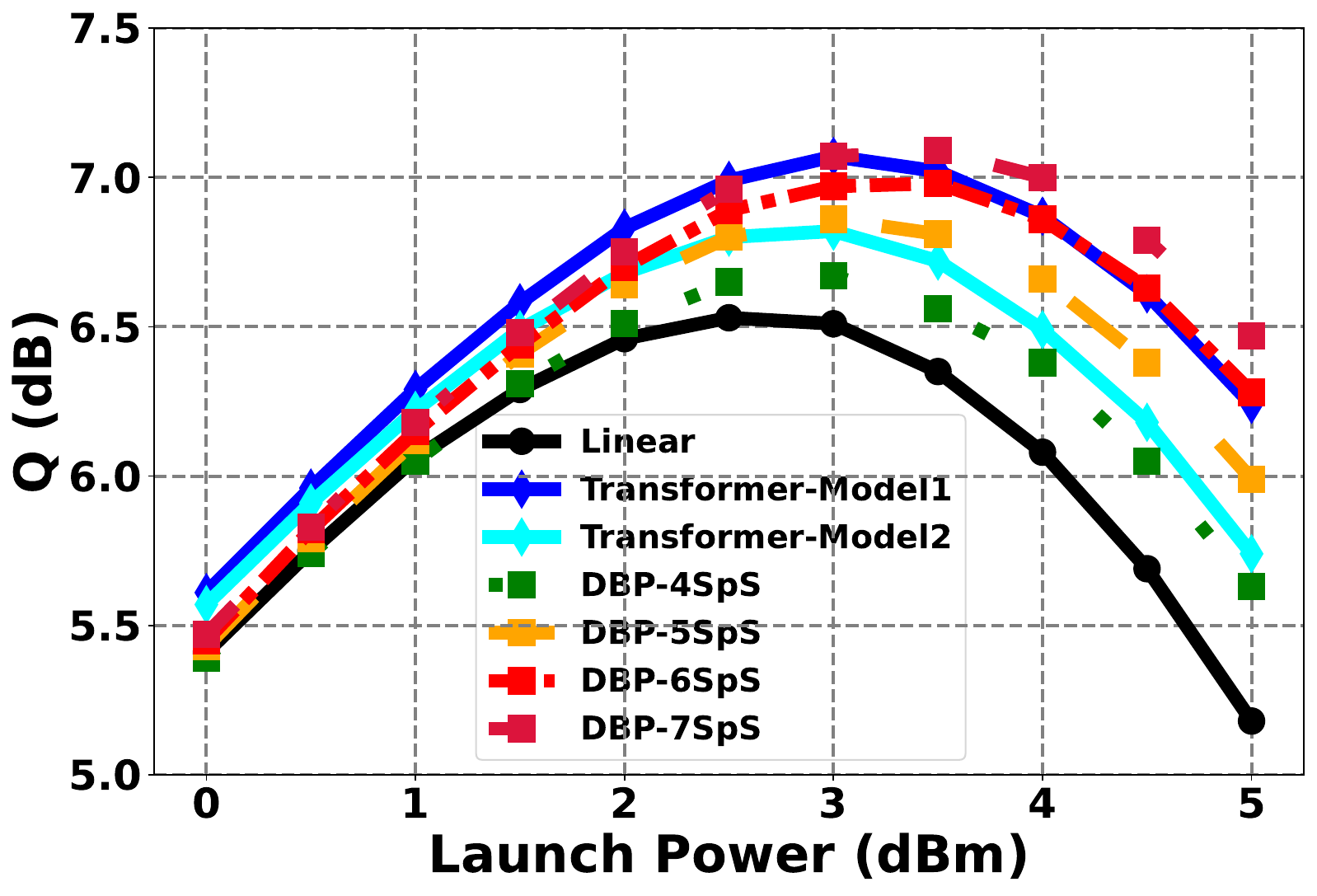}
		\caption{}
		\label{fig:transformerVsDbpPmd1d0K0d8Baud64GB}	
	\end{subfigure}
	\caption{Impact of increasing PMD: performance of Model1 and Model2 transformers of Table \ref{tbl:selectedmodelshpyerarams}  at 64GB compared with DBP with various steps-per-span at several launch powers. Figs (a), (b), and (c) show the results at 64 GBaud for PMD coefficient of 0.05, 0.50, and 1.0 $ps/\sqrt{km}$, respectively.}
	\label{fig:64GB_different_pmd}
\end{figure*}

Fig. \ref{fig:64GB_different_pmd} compares the performance of Model1 and Model2 at 64GBaud with DBP at three different PMD coefficient values, 0.05, 0.5 and 1.0 $ps/\sqrt{km}$, respectively. 
As depicted here, as the PMD coefficient increases, one need to employ DBP with higher steps-per-span to match the performance of both Transformer-NLC Models.

\begin{figure}[hbtp]
	\centering
	\includegraphics[width=0.5\linewidth]{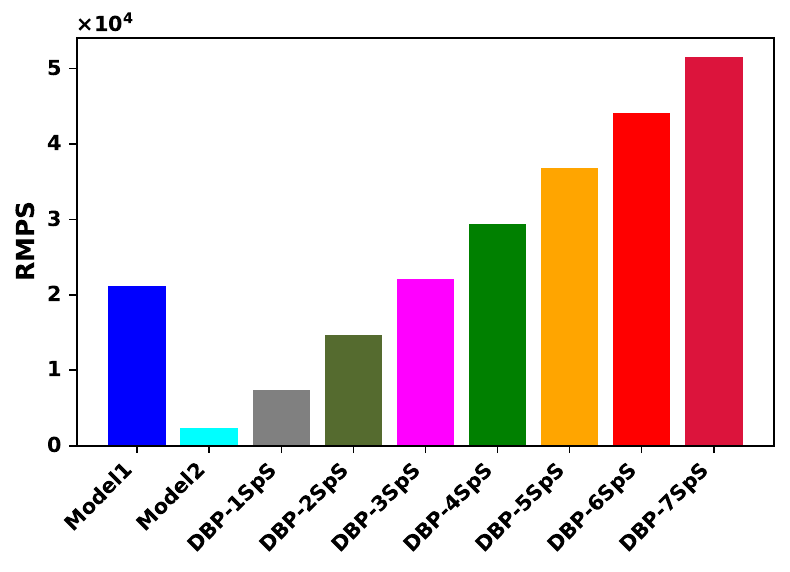}
	\caption{Complexity of Transformer-NLCs Model1 and Model2, and DBP with different steps-per-span}
	\label{fig:transformeranddbpcomplexity}
\end{figure}

Finally, to show a measure of complexity, Fig. \ref{fig:transformeranddbpcomplexity} shows the RMPS of Model1, Model2 and DBP with different steps-per-span. 
The complexity of DPB per symbol is computed as 
\begin{equation}
	C_{\text{DBP}} = 4 N_{span} N_{step} N_{up} \left( 2 \left( log_2(N_{FFT}) + 1\right)   + 1 \right) 
\end{equation}
where $N_{span}, N_{step}, N_{up}, \text{and } N_{FFT}$ denote the number of spans, the number of steps-per-span, the oversampling ratio, and the size of FFT (which is 1024 in this work), respectively.
As Fig. \ref{fig:transformeranddbpcomplexity} shows, the complexity of Model1 matches DBP with 3 steps-per-span while Model2 has the lowest complexity compared to any DBPs.

It is worthwhile to note that since in general higher baud rate signals are sent over fewer number of spans and also considering the impact from chromatic dispersion at different baud rates, the accumulated nonlinearity and the ability to train the models can vary significantly for different scenarios. 
Therefore, for practical applications, the ANN models need to be trained in several regions of operations. Then, transfer learning can be employed to fine-tune the models for each specific application  \cite{transfer_learn}.

\section{Conclusion}\label{TF_conc}
In this work, we introduced a new nonlinear channel  compensation method based on the self-attention mechanism in Transformers.
We show that Transformer-NLCs can be employed effectively for coherent long-haul transmissions.
Efficient embedding and detailed implementation of Transformers as well as the use of a physics-informed mask were presented. Moreover, the designs were optimized for block-processing in order to meet the practical high-throughput/low-latency requirements of long-haul optical transmission.
The presented designs were supported with numerical results for various transmission scenarios which show that one can use the Transformer-NLCs as an efficient, flexible, universal and parallelizable solution to compensate the fiber nonlinearity. 
To argue the benefits of Transformer-NLC, we also compared its performance versus complexity with DBP and LSTM-NLC. 
Specifically, through various changes in symbol rate and PMD coefficients, the advantage and flexibility of Transformer-NLC over DBP were demonstrated.
A direction for further studies can be simplifying the Transformer-NLC design. 
One possible approach is to modify the attention structure by designing a more efficient and kernel-based formulation of the self-attention. Furthermore, the impacts of quantization and pruning need to be studied for practical deployment through both post-training simplification and quantization-aware training.
%

\bibliographystyle{IEEEtran}
\bibliography{references}

\clearpage

\begin{center}
	\textbf{\LARGE Appendix}
\end{center}
\setcounter{section}{0}
\section{Attention Heat Maps}  \label{sec:appendix_heatmap}

\begin{figure}[]
	
	\begin{subfigure}{0.4\linewidth}
		\includegraphics[width=1.4\linewidth]
		{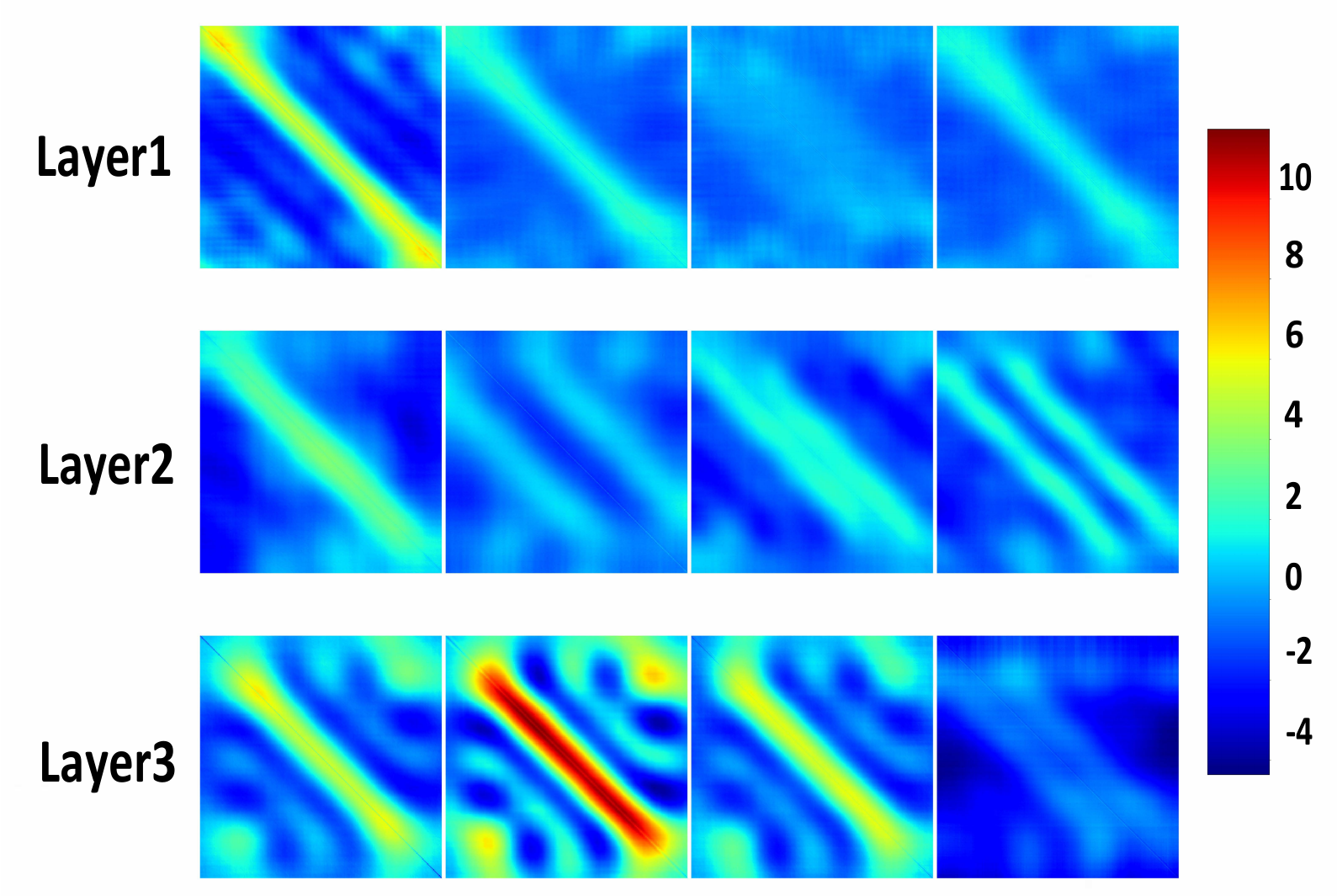}
		\caption{ }
		\label{fig:vanilaattenmatrix3layers4heads}
	\end{subfigure}
	\hfill
	\begin{subfigure}{0.25\linewidth}
		\centering
		\includegraphics[width=.6\linewidth]{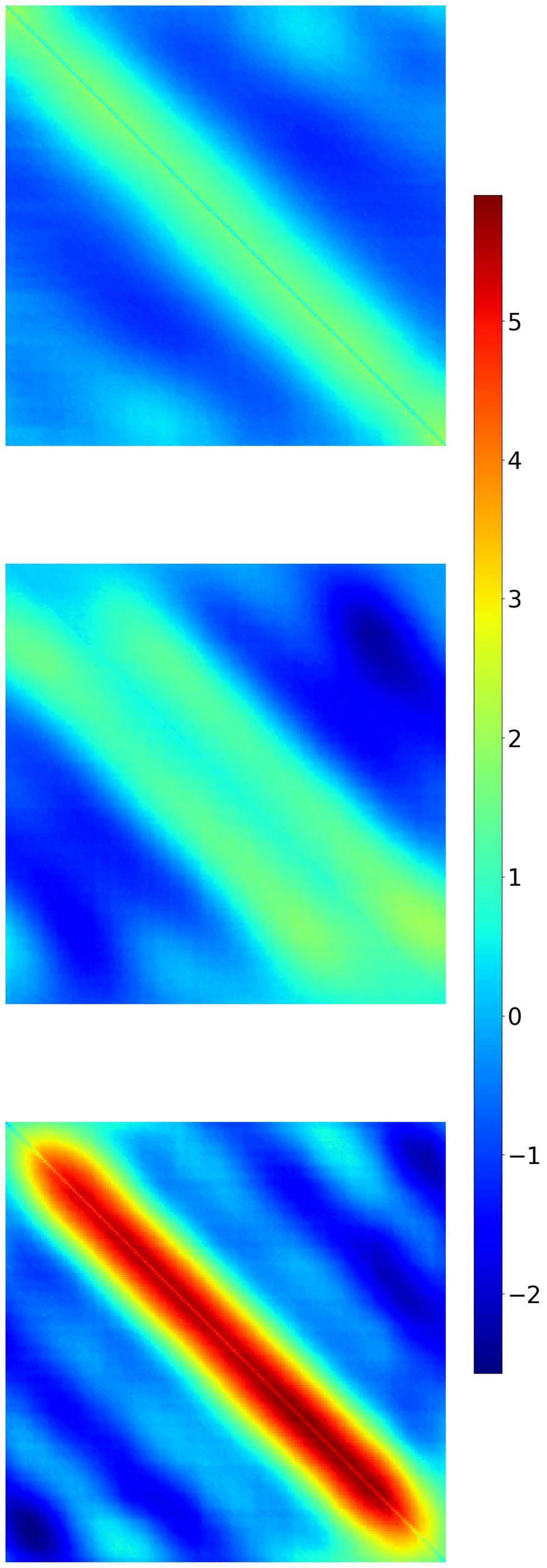}
		\caption{}		
		\label{fig:vanilaattenmatrix3layers4headswithmask}
	\end{subfigure}
	
	
	\begin{subfigure}{0.4\linewidth}
		\includegraphics[width=1.4\linewidth]{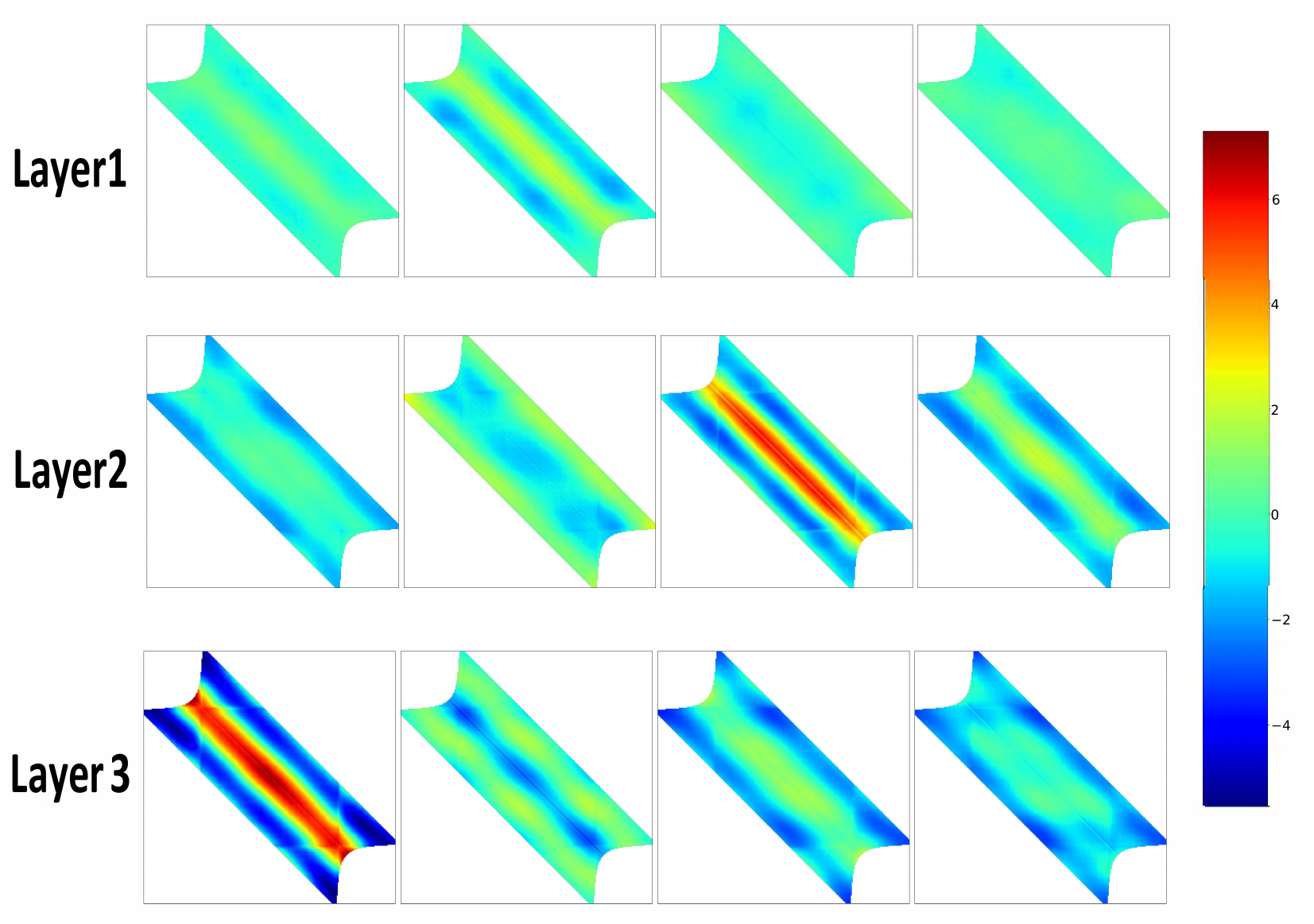}
		\caption{}
		\label{fig:attention_graph_withoutmask}		
	\end{subfigure}
	\hfill
	\begin{subfigure}{0.25\linewidth}
		\centering
		\includegraphics[width=0.7\linewidth]{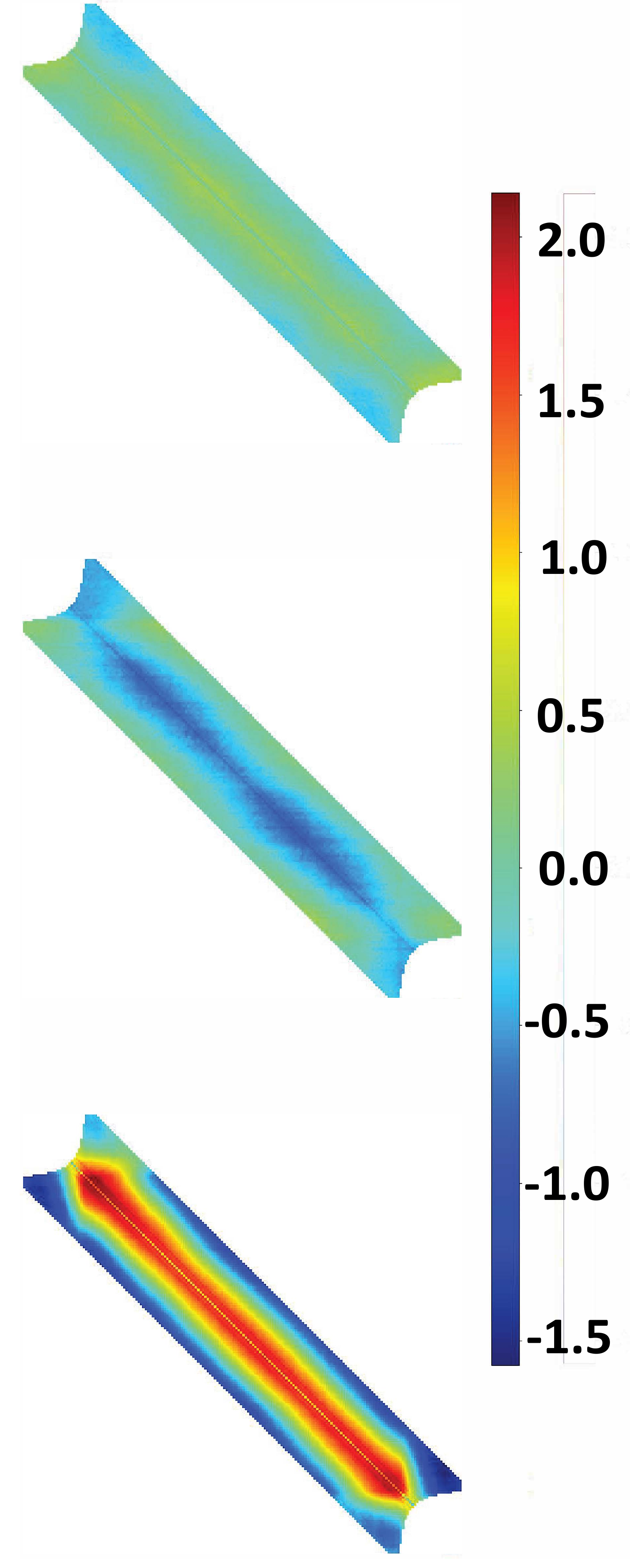}
		\caption{}
		\label{fig:attention_graph_withmask}
	\end{subfigure}	
	
	\caption{Heat maps for the attention matrices: 
		(a) shows the attention matrix for a Transformer with 3 encoder layers and 4 heads without any masks. Rows show the values for 3 layers and columns show the values for 4 heads. (b) shows the attention matrix for a Transformer with 3 encoder layers and one head. Rows show the values for 3 layers. (c) and (d) are the attention matrices with the proposed mask of $\rho$ equal to 2.6 for the examples in (a) and (b), respectively.}
\end{figure}

Here, we show several two-dimensional heat maps of $(QK^T)/\sqrt{d_k}$ (scaled multiplication of the queries and keys) with and without mask prior to the application of softmax function. 
Fig \ref{fig:vanilaattenmatrix3layers4heads} illustrates  $(QK^T)/\sqrt{d_k}$ values averaged over a batch for a Transformer with three encoder layers and four heads. 
As seen in here, the majority of large values of $(QK^T)/\sqrt{d_k}$ is around the main diagonal, which confirms that one can reduce the computational complexity by using the proposed masks. 
Thus, by using block masks we force some elements of attention matrix to zero which can reduce the computational complexity considerably. 
Fig.~\ref{fig:vanilaattenmatrix3layers4headswithmask} shows the $(QK^T)/\sqrt{d_k}$ values averaged over a batch for $\rho=2.6$.
Fig.~\ref{fig:attention_graph_withoutmask} and \ref{fig:attention_graph_withmask} shows another example of heat maps for a Transformer with three encoder layers and a single head. 

\include{appendix_b}

\end{document}

%% file: appendix_b.tex
\section{Impact of Model's Hyper-Parameters}\label{sec:TF_impact_params}
We present a deeper study on impact of different hyper-parameters on performance of the proposed Transformer-NLC. For brevity, we limit the scope of result presentation here to the 16QAM setup at 2 dBm launch power.
\subsection{Embedding Type}\label{subsec:embedding_type}

As mentioned in Section 3.2 of the manuscript,
we have explored two types of embedding-generator modules, one with MLPs and one with CNNs. Two configurations were used for the MLP: one MLP has only a linear layer which maps the input symbols ($R^{(2t+b) \times 4}$) to the embeddings ($R^{(\ell+b) \times d_{model}}$). The other one has two linear layers where the first one maps the input symbols ($R^{(2t+b) \times 4}$) to intermediate representations ($R^{(\ell+b) \times d_{interm}}$) and the second layer maps those to embeddings ($R^{(\ell+b) \times d_{model}}$). There is also an activation function (Leaky ReLU) between the two linear layers for the second MLP configuration. 
Note that over the input symbol sequence, the MLPs are applied to each position, similarly.
Furthermore, to employ the feature extraction capability of CNNs, a configuration is used for generating embeddings where we employ one convolutional layer with four input channel and $d_{model}$ output channels, a kernel size of nine and a stride of one. A Leaky ReLU activation is used after the convolutional layer.

Table~\ref{tbl:embedding_type} shows the impact of three different embedding-generator module types.
These results were obtained by a Transformer with the block size of 128, tap, key, hidden, and FFN hidden sizes of 64, and a window size of 7 for the window processing. 
As it can be seen from the table, CNN embedding generator module has a large positive impact on the model prediction capability with close RMPS values to the two MLP embedding generator modules. 

Another interesting observation comes from the implications of using masks: When a mask is used, the performance  with CNN generated embeddings did not decrease significantly in contrast to the  MLP generated embeddings. Similar trend was observed for other performance and complexity regions. 
Therefore, throughout the manuscript, CNN embedding generator modules were selected as the default choice in order to simplify the presentation.

\begin{table}[h!] 
	\centering
	\caption{Impact of the embedding type on performance for a Transformer with block size of 128, tap, key, hidden and FFN hidden sizes of 64, and a window size of 7.
	}
	\small
	\begin{tabular}{llllcllc}
		\Xhline{1pt} 
		& & & \multicolumn{2}{c}{\textbf{No Mask}} & \textbf{} & \multicolumn{2}{c}{\textbf{Mask with $\bm{\rho=2.6}$}} \\ \cline{4-5} \cline{7-8} 
		\textbf{Embedding} & \textbf{Q}& &	 $\mathbf{Q_{NN}}$    &  \textbf{RMPS (K)}   & \textbf{} & $\mathbf{Q_{NN}}$    & \textbf{RMPS (K)}  \\   \cline{1-2} \cline{4-5} \cline{7-8} 
		1 Layer MLP & 6.67 &  & 8.01 & 368 &  & 7.34 & 241 \\ \cline{1-2} \cline{4-5} \cline{7-8} 
		2 Layer MLP & 6.67 &  & 8.17 & 372 &  & 7.23 & 245 \\ \cline{1-2} \cline{4-5} \cline{7-8} 
		CNN & 6.67 &  & 8.72 & 354 &  & 8.70 & 222 \\  \Xhline{1pt} 
	\end{tabular}
	\label{tbl:embedding_type}
\end{table}

\clearpage
\newpage
\subsection{Block Size}\label{subsec:params_block_size}

\begin{figure}[htpb]
	\centering
	\includegraphics[width=0.7\linewidth]{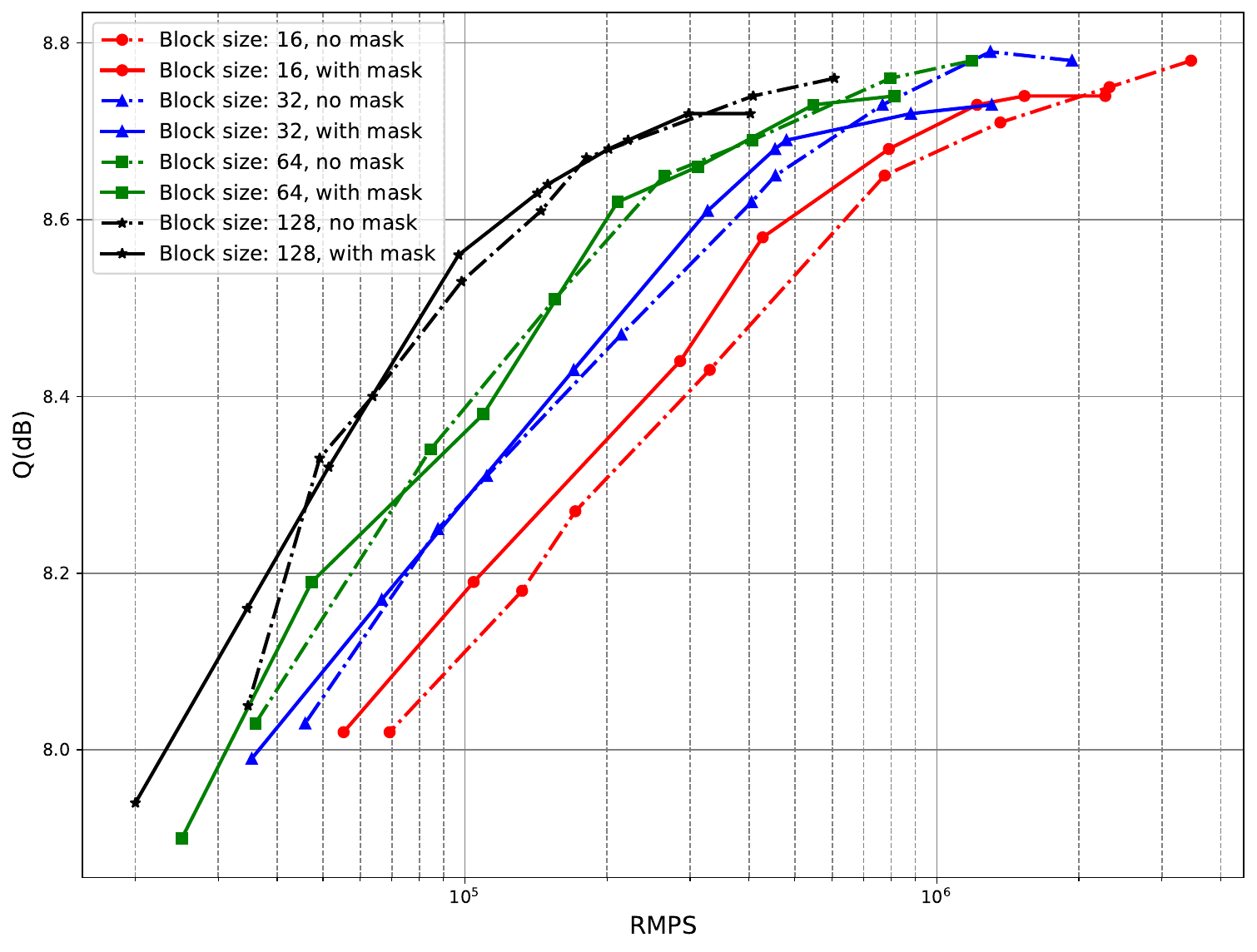}
	\caption{Impact of block size on the Transformer-NLCs.}
	\label{fig:tfblocksizeqvsflopenvelope}
\end{figure}

As we discussed in Eq. (5) of the manuscript, the computational complexity of a Transformer changes with the input block size. We have trained models with block sizes $b \in \{16, 32, 64, 128\}$ where Fig.~\ref{fig:tfblocksizeqvsflopenvelope} shows the envelopes of evaluated performance over different RMPS regions.
It can be seen here that as the block size increases, the performance envelopes shift to the lower complexity regions. We also do not observer a decrease or cap on the best performances.
This indicates that the block training approach is resilient and does not suffer by increasing the block size. Furthermore, it is apparent that the use of a mask is beneficial for all block sizes at almost all complexity regions except at the tail of high complexity region where extra symbols can provide minimal advantages specially at smaller block sizes.

For comparison, with an example we show the performance and complexity of Transformers with and without masks for the selected block sizes in Table~\ref{tbl:block_size_performance_complexity}. 
Note that the rest of the hyper-parameters are fixed in this case.
\begin{table}[htpb] 
	\centering
	\caption{Impact of block size on the Transformer-NLCs. Other hyper-parameters are as following: FFN hidden size = 64, hidden size = 64, key size = 64,window size = 15, number of heads = 4, tap size = 96.
	}
	\small
	\begin{tabular}{llllcllc}
		\Xhline{1pt} 
		& & & \multicolumn{2}{c}{\textbf{No Mask}} & \textbf{} & \multicolumn{2}{c}{\textbf{Mask with $\bm{\rho=2.6}$}} \\ \cline{4-5} \cline{7-8} 
		\textbf{block size} & \textbf{Q}& &	 $\mathbf{Q_{NN}}$    &  \textbf{RMPS (K)}   & \textbf{} & $\mathbf{Q_{NN}}$    & \textbf{RMPS (K)}  \\   \cline{1-2} \cline{4-5} \cline{7-8} 
		16 & 6.67 &  & 8.70 & 2039 &  & 8.73 & 1216 \\ \cline{1-2} \cline{4-5} \cline{7-8} 
		32 & 6.67 &  & 8.77 & 1141 &  & 8.70 & 711 \\ \cline{1-2} \cline{4-5} \cline{7-8} 
		64 & 6.67 &  & 8.71 & 708 &  & 8.69 & 451 \\ \cline{1-2} \cline{4-5} \cline{7-8} 
		128 & 6.67 &  & 8.71 & 511 &  & 8.68 & 307 \\  \Xhline{1pt} 
	\end{tabular}
	\label{tbl:block_size_performance_complexity}
\end{table}

By investigating Eq. (5) of the manuscript, we may expect that the complexity of a Transformer increases as we increase the block size.  However, the graph and table here show something different. To understand this, we analyze Eq. (5) of the manuscript for the cases where all the hyper-parameters are constant except the block size. We can rewrite this equation as follows:
\begin{equation}\label{key}
	RMPS_{att-blk} = \frac{c_1 (\ell+ b)^2}{b}
\end{equation}
where $c_1 = 2hd_k+3h$ and $N=\ell + b$. By expanding the above equation and after canceling out the common terms, we obtain
\begin{equation}\label{key}
	RMPS_{att-blk} = 2c_1\ell + \frac{c_1 \ell^2}{b} + c_1b
\end{equation}
where the first term is independent of the block size, the second term decreases as the block size increases (rectangular hyperbola) while the third term linearly increases as the block size increases. Therefore, at smaller bock sizes, the second term decreases more rapidly compared to the third term which is linear. However, at larger block sizes, the linear part (third term) grows lager than the rectangular hyperbola (second term) and as the block sizes increase, the overall complexity increases. 

\begin{figure}[htpb]
	\centering
	\includegraphics[width=0.7\linewidth]{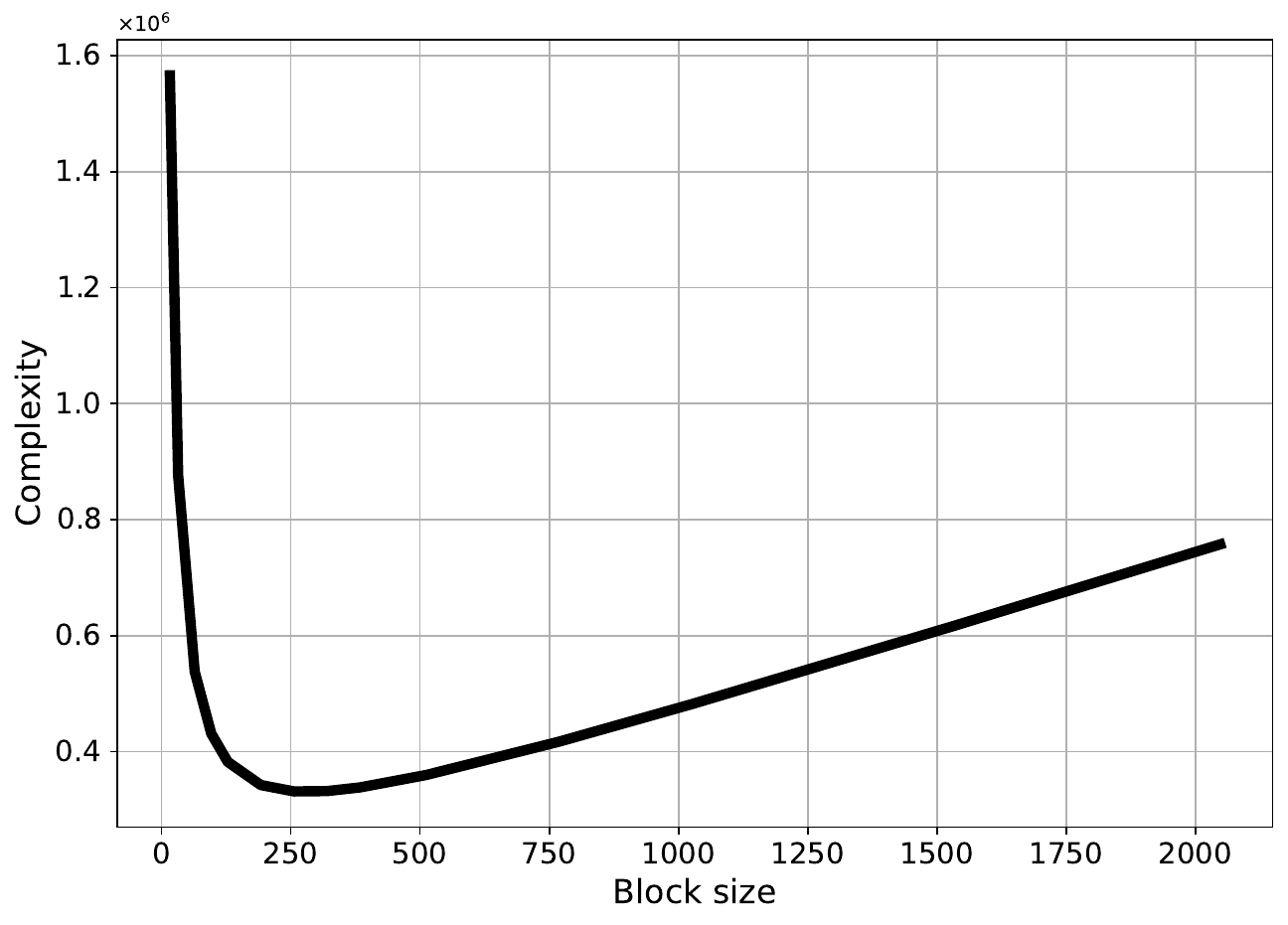}
	\caption{Complexity versus block size for a Transformer where all hyper-parameters are fixed except the block size. Tap size = 96, hidden size = 64, key size = 64, FFN hidden size = 64, number of heads = 4, and number of encoder layers = 2.}
	\label{fig:tfcomplexityvsblocksize}
\end{figure}

For illustration purpose, Fig.~\ref{fig:tfcomplexityvsblocksize} shows the complexity of a Transformer with respect to the block size, while all the other hyper-parameters are constant. The figure corroborates our claims by showing that as the block size is increased, the complexity reduces to a global minimum after which the linear term dominates and we see an increase in the complexity as the block size is increased further. 
It should also be noted that there are other types of Transformers where their attentions' computation complexities are of order of $O(Nd_{model})$ \cite{katharopoulos2020Trafsformers_R_RRN, zhai2021attention-free}. In those Transformers, the complexity should be reduced everywhere by increasing the block size.

\clearpage
\newpage
\subsection {Hidden Size}

\begin{figure}[h!]
	\centering
	\includegraphics[width=0.7\linewidth]{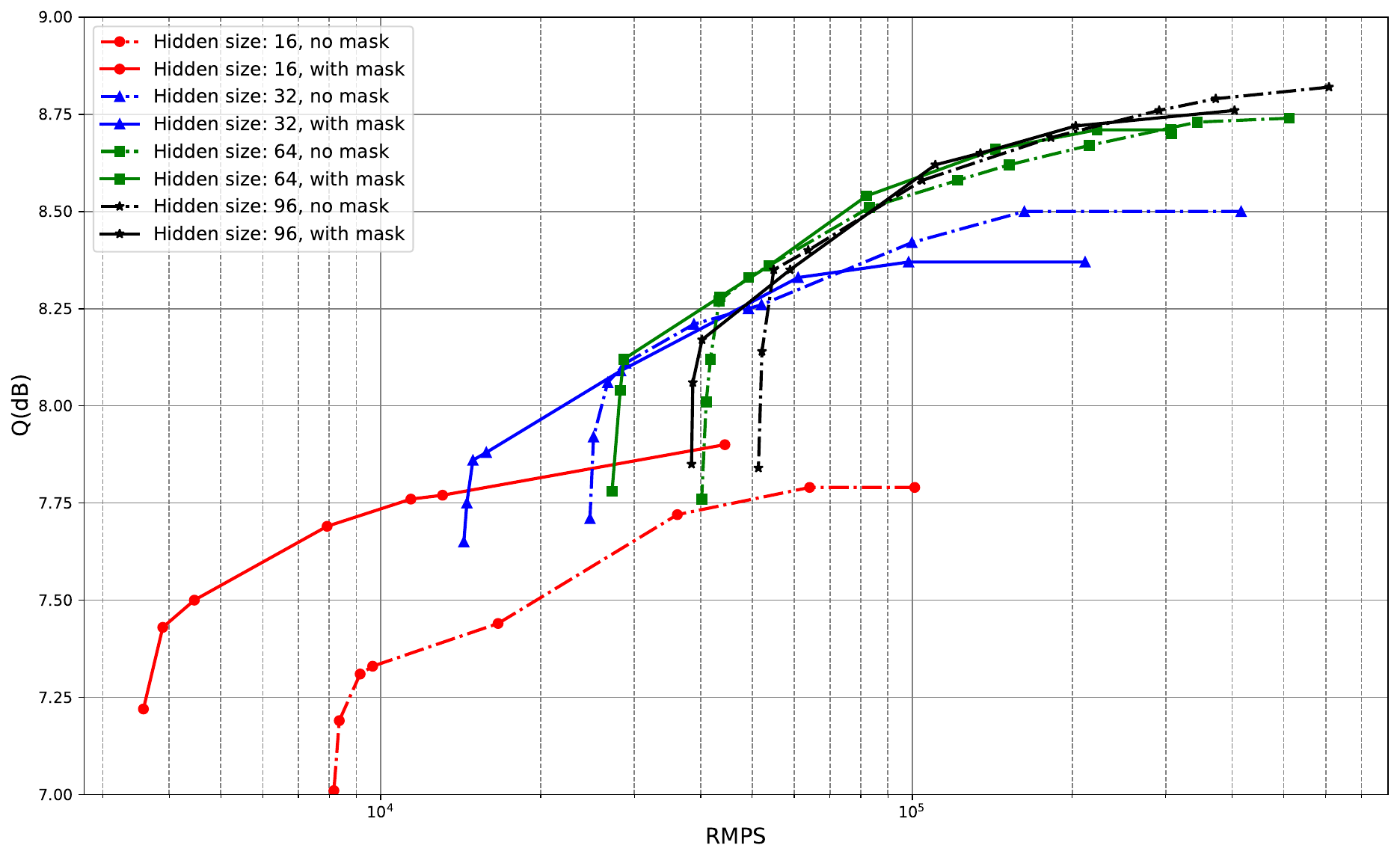}
	\caption{Impact of the hidden size on the Transformer-NLCs.}
	\label{fig:tfhiddensizeqvsflopbasedonrhoall3sets}
\end{figure}

The hidden size (the size of embeddings or $d_{model}$) impacts the Transformer's accuracy and computational complexity. A small hidden size may not be able to capture
the memory of nonlinear channel, although it decreases the computational complexity. On the other hand, a very large hidden size will inflate the complexity while may not necessarily improve the performance.
Figure~\ref{fig:tfhiddensizeqvsflopbasedonrhoall3sets} shows the effect of hidden size on performance versus complexity.
As it can be seen, at lower complexities, there is a loss on the envelope of performance when the mask is not used to optimize the complexity for limited available resources. However, at the higher complexity regions, the required hidden size is saturated and there is no gap related to use of mask on the envelope of performance. 
For the selected setup, to maximize the performance, hidden size of 64 is the best option overall which covers a wide range of complexities while giving the best performance.

\clearpage
\newpage
\subsection {Tap Size}
\begin{figure}[h]
	\centering
	\includegraphics[width=0.7\linewidth]{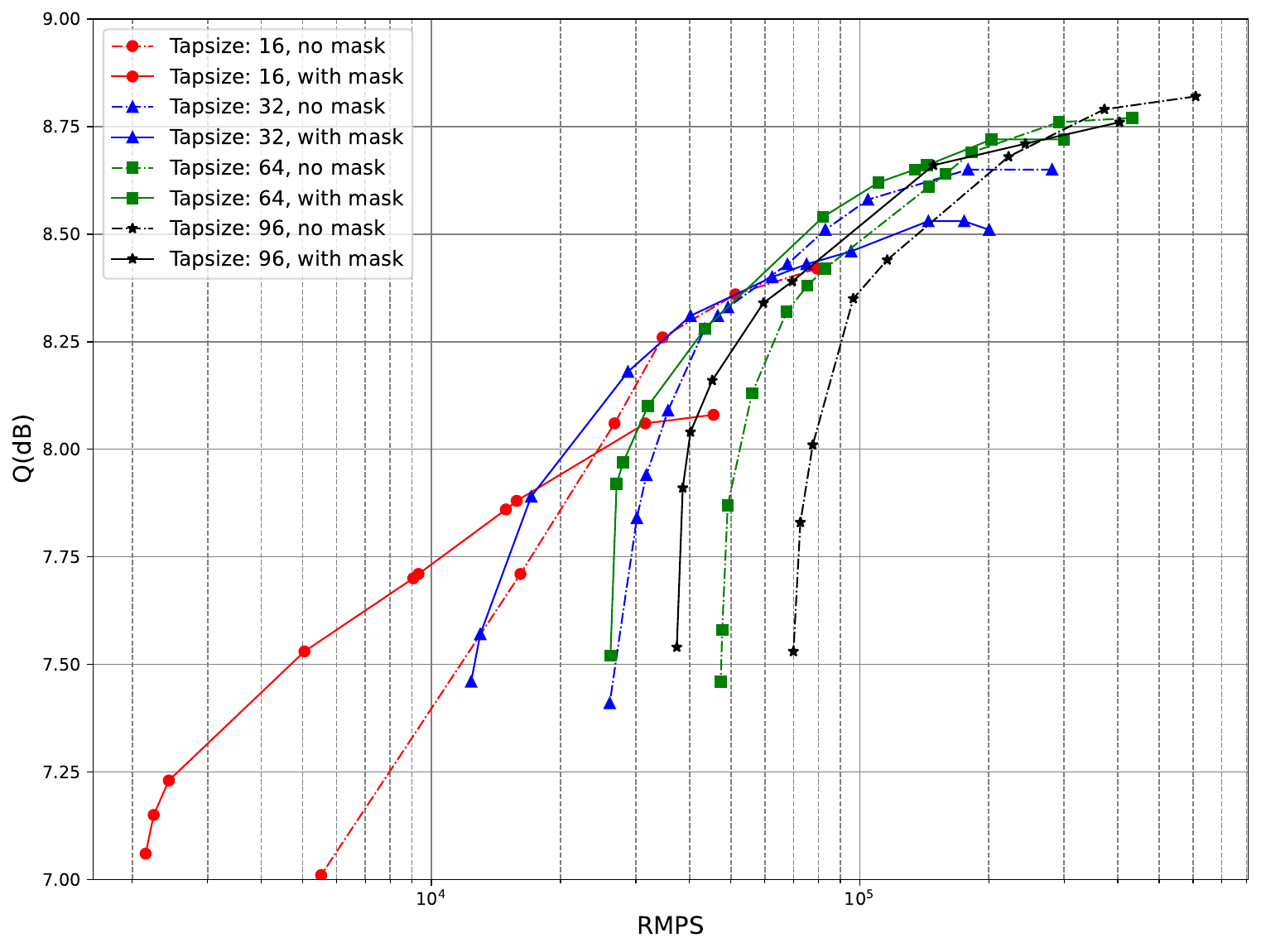}
	\caption{Impact of the tap size on the Transformer-NLCs.}
	\label{fig:tftapsizeqvsflopbasedonrhoall3sets}
\end{figure}

As we discussed in Section 3 of the manuscript, we need to provide the surrounding input symbols to the model in order to compute the nonlinear interference in form of $E_{XI}$ and $E_{XQ}$ for a target symbol at the output of Transformer. This is mainly due to the channel memory from the accumulated dispersion during fiber propagation. The number of these surrounding symbols are determined by the tap size. 
We trained models with various tap sizes where the results are depicted in Fig.~\ref{fig:tftapsizeqvsflopbasedonrhoall3sets}.
As it can be seen here, for higher complexities, the tap sizes of 64 and 96 provide better performances while among them, the tap size of 64 covers a wider range of complexities.
Furthermore, with a large enough tap size, the use of a mask has almost no impact at the higher performances while provides complexity advantage on the lower side of the performance envelope.

Note that for each taps size, the performance envelopes for Transformers with and without mask demonstrate a crossing complexity threshold. Below that threshold, the use of a mask provides superior performance while for models with higher complexity the use of mask demonstrates some performance loss especially at lower tap sizes.
This loss can be attributed to the extended interactions across neighboring symbols through the block-processing structure where the symbols in the middle of the block still have connections to a larger than defined tap size inside attention calculation. However, this extended tap size is blocked by the use of a mask. Therefore, for a given small tap size at higher complexity region, it is better to avoid using a mask to be able to access information from more neighboring symbols in the block-processing approach.

Generally at lower complexity regions, one should reduce the tap size in order to achieve better performances as seen by performance curves associated to 16 and 32 tap sizes.
Also note that according to envelope of performance versus complexity, one can still get a better performance with a masked structure if a proper tap size is selected. 
In other words, the best performance curve across all tap sizes is defined by the masked models.

\clearpage
\newpage
\subsection{Number of Layers}
\begin{figure}[htbp]
	\centering
	\includegraphics[width=.7\textwidth]
	{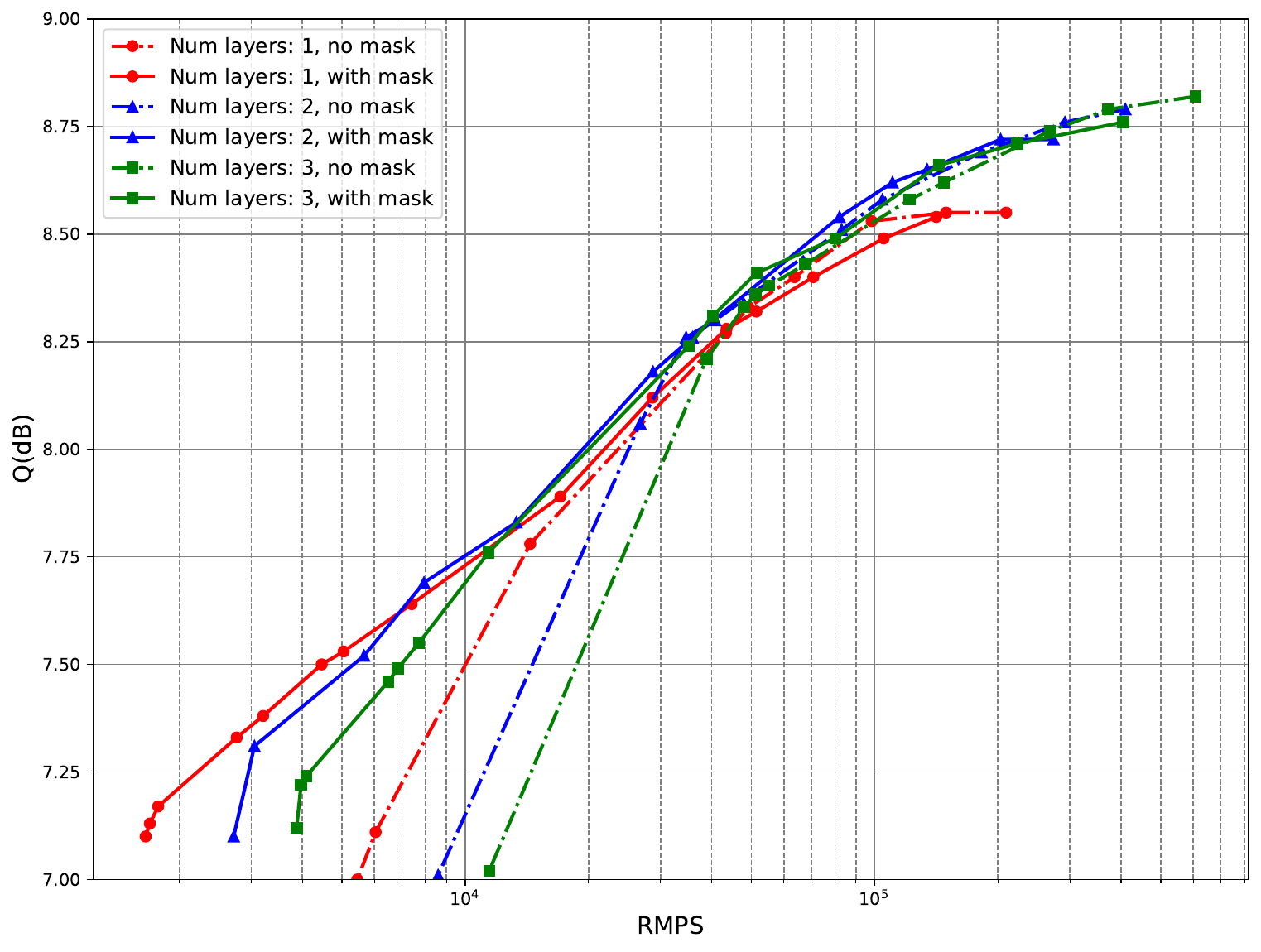}
	\caption{Impact of the number of encoder layers on the Transformer-NLCs.}
	\label{fig:tfnumencoderlayersqvsflopbasedonrhoall3sets}
\end{figure}

As discussed in Section 2.2 and depicted in Fig. 2 of the manuscript, 
core of Transformers can be stacked in several layers which linearly scales the computational complexity provided that the other parameters are unchanged.
In order to see the impact of number of layers in the Transformer-NLCs, we ran simulations with one to three layers and explored the performance versus complexity trade-off for each design. 
The results are depicted in Fig.~\ref{fig:tfnumencoderlayersqvsflopbasedonrhoall3sets}. 
One can conclude from this figure that Transformers with three layers of encoder perform slightly better compared to the ones with two layers of encoder at high complexity regions. However, for a wide range of complexities, the two-layer configuration is the best model to achieve the performance envelope.
Moreover, the use of a mask generally improves the performance at lower complexity regions and extends the savings in resources as we face the lower complexity limits.     

\clearpage
\newpage
\subsection{Number of Heads}
\begin{figure}[htbp]
	\centering
	\includegraphics[width=.7\textwidth]
	{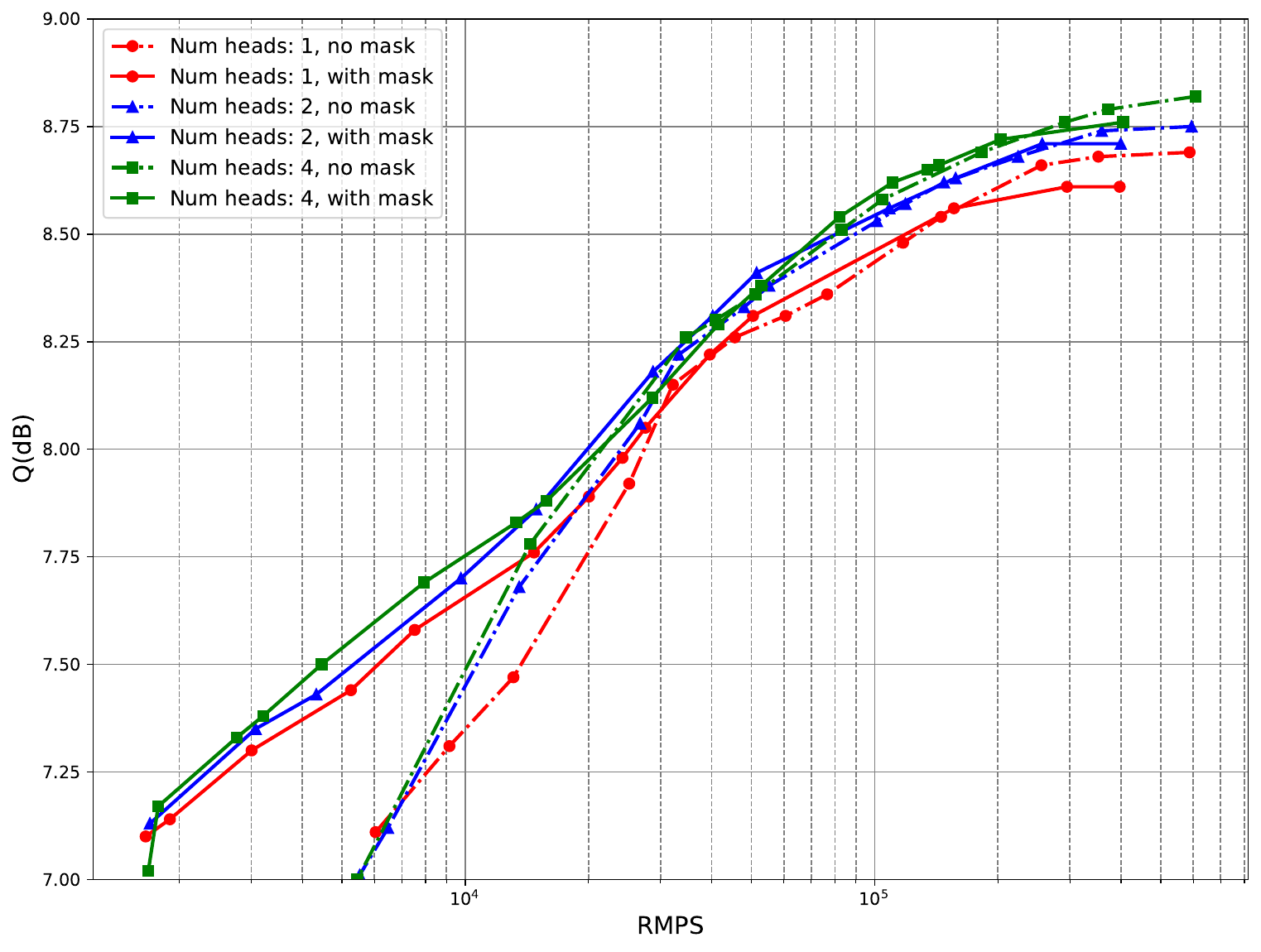}
	\caption{Impact of the number of heads on the Transformer-NLCs.}
	\label{fig:tfnumheadsqvsflopbasedonrhoall3sets}
\end{figure}

The impact of number of heads in the multi-head attention structure is studied next where we have tested one, two, and four heads. Note that in general, a higher number of heads increases the parallelization capabilities of the Transformer while slightly improves the performance. 
Results for the selected cases are depicted in Fig.~\ref{fig:tfnumheadsqvsflopbasedonrhoall3sets}. 
This figure shows that for a wide range of complexities, a higher number of attention heads results in slightly better performance for the same complexity. Also as observed before, the use of a mask has no significant impact on the overall conclusion while it improves performance at lower complexity regions.

\clearpage
\newpage
\subsection{Number of Neighbors at the Output Layer}
\begin{figure}[h!]
	\centering
	\includegraphics[width=0.7\linewidth]{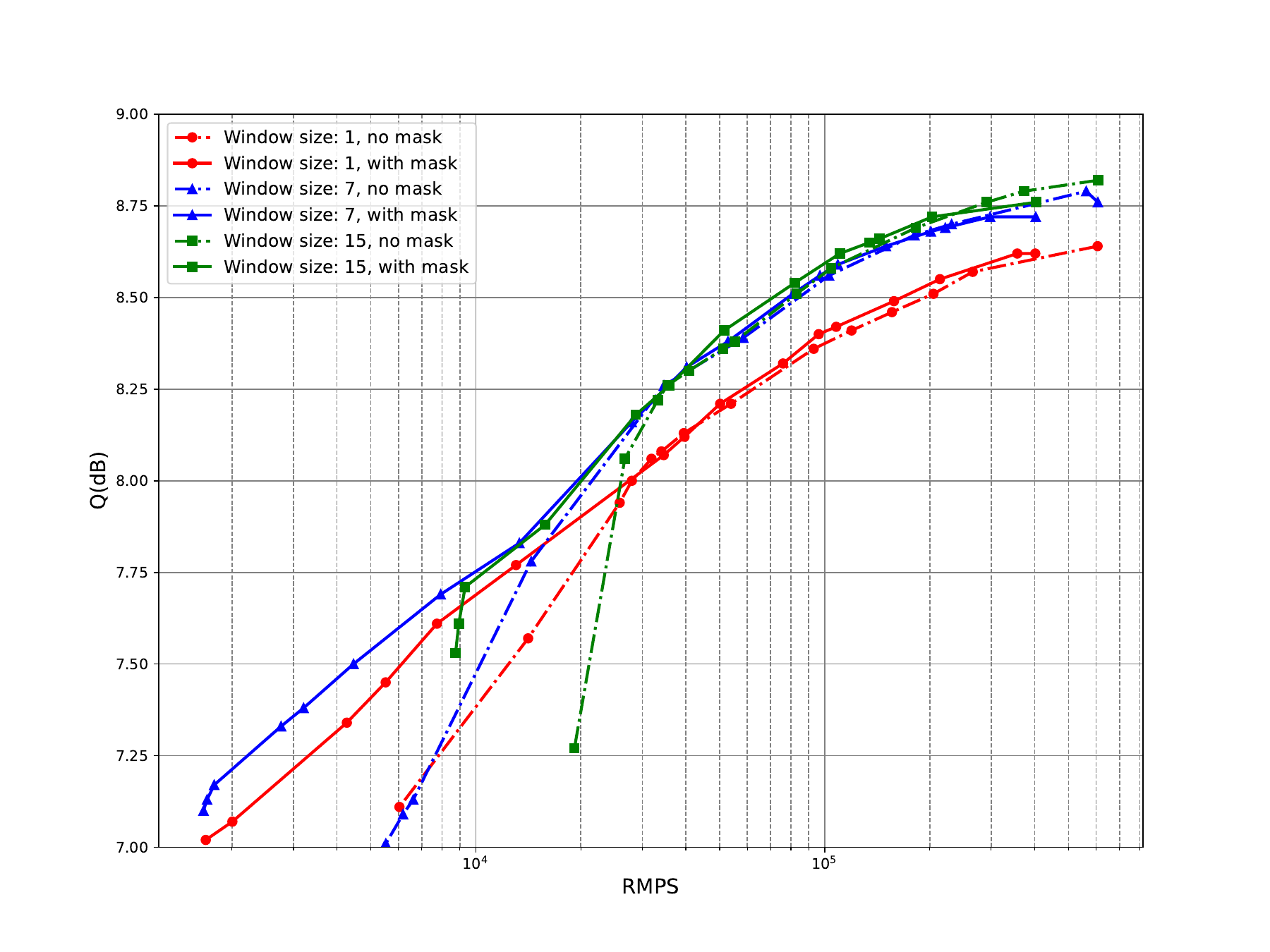}
	\caption{Impact of the window size at the output layer on the Transformer-NLCs.}
	\label{fig:tfneighborsqvsflopbasedonrhoall3sets}
\end{figure}

Finally, we explore the impact of number of neighbors, $2w$, or equivalently the window size $W=2w+1$ on the performance and complexity of Transformers as explained in a Section 3.4 of the manuscript.
Results are depicted in Fig.~\ref{fig:tfneighborsqvsflopbasedonrhoall3sets}. 
As the figure shows, increasing the window size can provide a gain in performance of up to 0.2 dB over a wide range of complexities. Note that a window size of 7 seems to be optimal for mid and high complexity regions. 
It can also be noted that increasing the window size is not beneficial especially at lower complexity regions where the extra information on the last layer is not helpful if the core of Transformer is limited in resources.
Also, the use of a mask follows the general conclusion that a mask can improve the performance at lower complexities.

